\documentclass{aa}
\usepackage{amsmath}
\usepackage{graphicx}
\usepackage{txfonts}

\begin{document}

   \title{Planetary formation and water delivery in the habitable zone \\
          around solar-type stars in different dynamical environments}
   \author{P. S. Zain\inst{1,2}\thanks{pzain@fcaglp.unlp.edu.ar}, 
           G. C. de El\'{\i}a\inst{1,2}, 
           M. P. Ronco\inst{1,2},
           \and O. M. Guilera\inst{1,2}
          }

   \offprints{Patricio Salv. Zain
    }
  \institute{Instituto de Astrof\'{\i}sica de La Plata, CCT La Plata-CONICET-UNLP \\                                                                                       
   Paseo del Bosque S/N (1900), La Plata, Argentina                                                                                                                         
   \and Facultad de Ciencias Astron\'omicas y Geof\'{\i}sicas, Universidad Nacional de La Plata \\                                                                           
   Paseo del Bosque S/N (1900), La Plata, Argentina 
                }

   \date{Received / Accepted}


\abstract
{Observational and theoretical studies suggest that there are
many and various planetary systems in the Universe.}
{We study the formation and water delivery of planets in the habitable zone (HZ) around solar-type stars. In particular, we study different dynamical environments that are defined by the most massive body in the system.}
{First of all, a semi-analytical model was used to define the mass of the protoplanetary disks that produce each of the five dynamical scenarios of our research. Then, we made use of the same semi-analytical model to describe the evolution of embryos and planetesimals during the gaseous phase. Finally, we carried out $N$-body simulations of planetary accretion in order to analyze the formation and water delivery of planets in the HZ in the different dynamical environments.}
{Water worlds are efficiently formed in the HZ in different dynamical scenarios. In systems with a giant planet analog to Jupiter or Saturn around the snow line, super-Earths tend to migrate into the HZ from outside the snow line as a result of  interactions with other embryos and accrete water only during the gaseous phase. In systems without giant planets, Earths and super-Earths with high water by mass contents can either be formed in situ in the HZ or migrate into it from outer regions, and water can be accreted during the gaseous phase and in collisions with water-rich embryos and planetesimals.}
{The formation of planets in the HZ with very high water by mass contents seems to be a common process around Sun-like stars. Our research suggests that such planets are still very efficiently produced in different dynamical environments. Moreover, our study indicates that the formation of planets in the HZ with masses and water contents similar to those of  Earth seems to be a rare process around solar-type stars in the systems under consideration.}

\authorrunning{P. S. Zain et al.}
\titlerunning{Planetary formation and water delivery in the habitable zone in different dynamical environments}

\maketitle
\section{Introduction}

In the past years, huge progress has been made in planetary sciences thanks to the increasing power of numerical simulations and the refinement of detection techniques. The number of confirmed exoplanets to date is 3603 (http://exoplanet.eu), and it continues to grow as we speak. Moreover, several observational works \citep{Cumming2008,Howard2013} and theoretical studies \citep{Mordasini2009,Ida2013} have suggested a wide diversity of planetary architectures in the Universe. From these considerations, a detailed analysis concerning the potential habitability of the terrestrial-like planets formed in such systems needs to be developed in order to determine their astrobiological interest. 

From a theoretical point of view, it is difficult to determine which planets could be able to harbor life. In general terms, the most important condition required for a planet to be habitable is the permanent presence of liquid water on its surface. The circumstellar region inside which a planet could retain liquid water on its surface is called habitable zone (HZ). However, it is worth remarking that a planet located in the HZ is not necessary habitable. In fact, the maintenance of conditions of habitability on a planet require satisfying other points that
are related to suitable atmospheric properties, organic material, magnetic field, and plate tectonics that replenish the atmosphere with CO$_{2}$, among others.

Many works based on $N$-body simulations have been developed in order to analyze the formation of terrestrial-like planets and water delivery in the HZ in distinct dynamical environments around stars of different spectral types. Several of these works considered systems without giant planets. \citet{Raymond2007} studied the habitable planet formation around stars with masses between 0.2$M_{\odot}$ and 1$M_{\odot}$. \citet{Ogihara2009} also studied planetary accretion from planetesimals inside the snow line around M stars, including tidal interactions with the disk. More recently, \citet{deElia2013} analyzed the formation of terrestrial planets in high-mass disks without gas giants around solar-type stars, while \citet{Ronco2014} studied the diversity of planetary systems that might form around Sun-like stars in low-mass disks in absence of giant planets. In a similar way, \citet{Dugaro2016} analyzed the formation of terrestrial-like planets and water delivery in the HZ in systems without gaseous giants around M0- and M3-type stars, while \citet{AlibertBenz2017} studied the formation and composition of terrestrial planets around stars with mass 0.1 $M_{\odot}$. Moreover, \citet{Cielsa2015} studied planetary formation around stars of different masses considering water-mass fraction of bodies beyond the snow line similar to comets, while \citet{Mulders2015} explored how water delivery is affected by dispersions of the snow line location.

Several works have also studied the planetary formation process and water delivery in the HZ in systems harboring at least one giant planet in different dynamical configurations. On the one hand, \citet{Mandell2007} explored the formation of terrestrial-like planets in the HZ during and after giant planet migration around Sun-like stars. These authors proposed models with a single migrating giant planet, as well as with one inner migrating and one outer non-migrating giant planet. In a similar way, \citet{Fogg2009} studied the formation and evolution of terrestrial-like planets around solar-type stars in systems in which a gaseous giant halts its migration at semimajor axes in the range 0.13 au - 1.7 au through the dissipation of the gas disk. On the other hand, \citet{Raymond2004} and \citet{Raymond2006} examined the accretion process of terrestrial planets and water delivery in the HZ around 1$M_{\odot}$ under the effects of a Jovian planet in the outer disk, while \citet{Raymond2011} analyzed the habitable planet formation in planetary systems that contain multiple marginally unstable gas giants. It is worth remarking that all these studies assumed that water is delivered to planets via collisions with volatile-rich bodies that condensed past the snow line, beyond about 3 au. 

In this paper, we present results of numerical simulations aimed at studying the processes of formation of terrestrial planets and water delivery in the HZ around solar-type stars in different planetary environments. In particular, we propose to analyze these processes in different systems that harbor a planet analog to Jupiter, Saturn, Neptune, and super-Earths of 5 $M_{\oplus}$ and 2.5 $M_{\oplus}$ around the snow line at the end of the gaseous phase.  
To do this, we first make use of a semi-analytical model to calculate the formation of the systems during the gaseous phase, obtaining initial conditions to carry out $N$-body simulations of planetary accretion. It is important to remark that to perform $N$-body simulations aimed at analyzing the last stage of planetary accretion after the dissipation of the gas disk, it is necessary to define physical and orbital initial conditions for the planets and planetesimals of the system. Most of the works that studied this topic considered arbitrary initial conditions to study the post-oligarchic planet formation growth \citep{Raymond2004,OBrien2006,Mandell2007,Fogg2007,Raymond2009,Ronco2014}. However, \citet{Ronco2015} showed that more realistic initial conditions, obtained by a planet formation model during the gaseous phase, lead to different accretion histories of the surviving planets even though the global properties of the planetary systems remain similar.

This paper is therefore structured as follows. The general properties of the protoplanetary disks used in our study are presented in Section 2. In Section 3 we present the semi-analytical model that allows us to describe the evolution of the systems during the gaseous phase. Then, the $N$-body code used to carry out our dynamical simulations is presented in Section 4. In Section 5 we describe our results and carry out a detailed analysis of all simulations. Finally, discussions and conclusions are presented in Section 6.

\section{Properties of the protoplanetary disk}

One relevant parameter that determines the distribution of material in a protoplanetary disk is the surface density.
The surface density profile adopted in our model of a protoplanetary disk is based on the evolution of a thin Keplerian disk that
is subject to the gravity of a
point-mass central star $M_{\star}$ \citep{Lynden-Bell1974,Hartmann1998}. Thus, the gas-surface density profile $\Sigma_{\text{g}}(R)$ is given by
\begin{equation}
\Sigma_{\text{g}}(R) = \Sigma_{\text{g}}^{0}\left(\frac{R}{R_{\text{c}}}\right)^{-\gamma} \text{exp}\left[-\left(\frac{R}{R_{\text{c}}}\right)^{2-\gamma}\right],
\label{e1}
\end{equation}
where $R$ is the radial coordinate in the midplane, $\Sigma_{\text{g}}^{0}$ a normalization constant, $R_{\text{c}}$ the characteristic radius, and $\gamma$ the exponent that defines the surface density gradient. It is worth noting that Eq.~\ref{e1} is an analytic solution to a simplified model for a viscous disk with a particular viscosity law. Real disks are not guaranteed to follow this profile. When we integrate Eq.~\ref{e1} over the total disk area, $\Sigma_{\text{g}}^{0}$ is written in terms of the total disk mass $M_{\text{d}}$ by
\begin{equation}
\Sigma_{\text{g}}^{0} = (2 - \gamma)\frac{M_{\text{d}}}{2 \pi R_{\text{c}}^{2}}\end{equation}  
for $\gamma \neq$ 2.

In the same way, the solid-surface density profile $\Sigma_{\text{s}}(R)$ is given by
\begin{equation}
\Sigma_{\text{s}}(R) = \Sigma_{\text{s}}^{0}\eta_{\text{ice}}\left(\frac{R}{R_{\text{c}}}\right)^{-\gamma} \text{exp}\left[-\left(\frac{R}{R_{\text{c}}}\right)^{2-\gamma}\right],
\label{e3}
\end{equation} 
where $\Sigma_{\text{s}}^{0}$ is a normalization constant, and $\eta_{\text{ice}}$ the parameter that represents an increase in the amount of solid material that is due to the condensation of water beyond the snow line. According to \citet{Hayashi1981}, $\eta_{\text{ice}}$ adopts values of 0.25 and 1 inside and outside the snow line, respectively, which is located at 2.7 au. However, the results derived by \citet{Lodders2003} and \citet{Lodders2009} imply a jump in the surface density of solids at the snow line that is substantial, but smaller than the factor of 4 implied by the model of \citet{Hayashi1981}. We therefore here use values for $\eta_{\text{ice}}$ of 0.5 and 1 inside and outside the snow line (2.7 au), respectively, which is in agreement with what
has been proposed by \citet{Lodders2003} and \citet{Lodders2009}.

The relation between the gas and solid surface densities is linked with the star metallicity $[\text{Fe}/\text{H}]$ by 
\begin{equation}
\left(\frac{\Sigma_{\text{s}}^{0}}{\Sigma_{\text{g}}^{0}}\right)_{\star} = \left(\frac{\Sigma_{\text{s}}^{0}}{\Sigma_{\text{g}}^{0}}\right)_{\odot}10^{[\text{Fe}/\text{H}]}
= z_{0}10^{[\text{Fe}/\text{H}]},
\label{e4}
\end{equation} 
where $z_{0}$ is the primordial abundance of heavy elements in the Sun and has a value of 0.0153 \citep{Lodders2009}. 
 
Several parameters, such as $M_{\star}$, $[\text{Fe}/\text{H}]$, $\gamma$, $R_{\text{c}}$, and $M_{\text{d}}$, must be quantified to specify the scenario of our simulations. On the one hand, our simulations assume a central star of 1 $M_{\odot}$ and solar metallicity (namely, $[\text{Fe}/\text{H}] = 0$). On the other hand, $\gamma$ and $R_{\text{c}}$ are assumed to be 0.9 and 25 au, respectively, which are in agreement with the values derived by \citet{Andrews2010} from the analysis of 17 protoplanetary disks in the 1 Myr old Ophiuchus star-forming region.

To define the mass $M_{\text{d}}$ of the protoplanetary disks used in our simulations, it is necessary to study the evolution of the systems during the gas phase in detail. In the present research, we aim to analyze the evolution of planetary systems in disks that leads to the formation of a planet analog to Jupiter, Saturn, Neptune, and a super-Earth around the snow line. To specify which type of protoplanetary disks can lead to these scenarios, we make use of a semi-analytical model that is able to analyze the evolution of a planetary system in the gaseous phase. As we show in the next sections, this model allows us to define the mass $M_{\text{d}}$ of the protoplanetary disks and the initial conditions for the distribution of embryos and planetesimals that need to be used in the $N$-body simulations.

\section{Semi-analytical model: gaseous phase}

\subsection{General description}

\begin{figure}[t]
 \centering
 \includegraphics[angle=270, width= 0.5\textwidth]{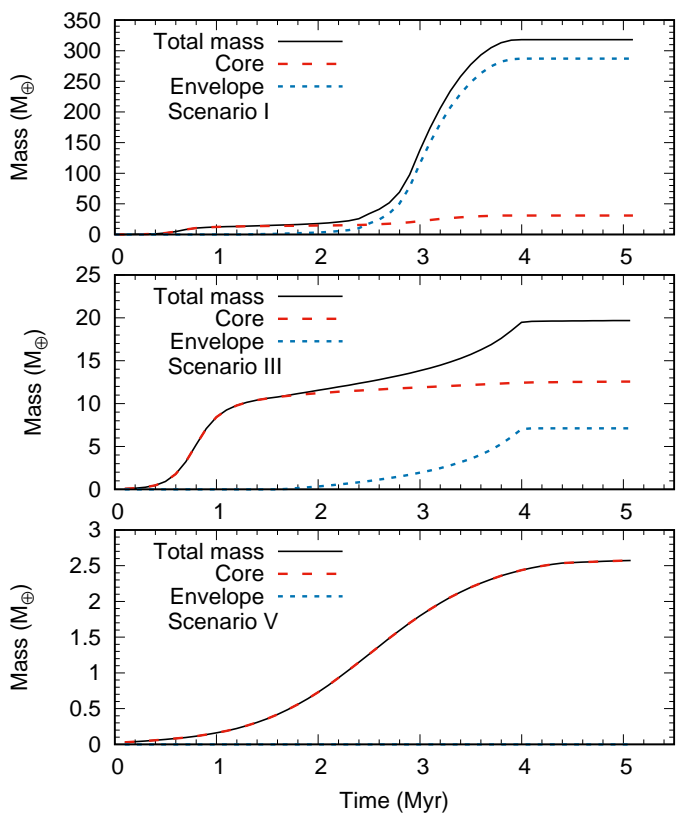}    
 \caption{Time evolution of the total mass, core mass, and envelope mass of a Jupiter-like planet (top), a Neptune-like planet (center), and a super-Earth of $2.5M_\oplus$ (bottom), which are the most massive planets formed around the snow line in their corresponding planetary systems. A color figure is available in the electronic version.}
\label{fig:Cores-y-Envolturas}                                                                                                
\end{figure}

The model of planet formation considered in this work has been described in \citet{Guilera2010,Guilera2014}, and we added some minor improvements. This model calculates the formation of a planetary system immersed in a protoplanetary disk that evolves in time. The disk is characterized by a gaseous component and a population of planetesimals. In our previous works, we considered that the gaseous component dissipated exponentially in time. However, in the present work, we implemented the evolution of the gaseous component that is due to a viscous accretion disk \citep{Pringle1981} with photoevaporation, modeled by
\begin{eqnarray}
  \frac{\partial \Sigma_{\text{g}}}{\partial t}= \frac{3}{R}\frac{\partial}{\partial R} \left[ R^{1/2} \frac{\partial}{\partial R} \left( \nu \Sigma_{\text{g}} R^{1/2}  \right) \right] + \dot{\Sigma}_{\text{w}}(R), 
\label{eq:eq1-sec3}
\end{eqnarray}
where $\nu= \alpha c_{\text{s}} \text{H}_{\text{g}}$ is the viscosity \citep{ShakuraSunyaev1973}, where $\alpha$ is a parameter, $c_{\text{s}}$ is the sound speed, $\text{H}_{\text{g}}$ is the scale height of the disk, and $\dot{\Sigma}_{\text{w}}$ represents the sink term that is due to photoevaporation from the central star \citep{AlexanderClarke2006,AlexanderArmitage2007}. 

The population of planetesimals can be modeled with an advection equation due to the conservation of mass given by
\begin{eqnarray}
  \frac{\partial \Sigma_{\text{p}}}{\partial t} - \frac{1}{R}\frac{\partial}{\partial R} \bigg(Rv_{\text{mig}}\Sigma_{\text{p}}\bigg) = \mathcal{F}(R), 
\label{eq:eq2-sec3}
\end{eqnarray}
where $v_{\text{mig}}$ is the planetesimal migration velocity, and $\mathcal{F}$ represents the sink terms that are due to the accretion by the embryos. We did not consider the planetesimal collisional evolution here. The evolution of the eccentricities and inclinations of the planetesimals is governed by the gravitational stirring produced by the embryos \citep{Ohtsuki2002} and by the damping that is due to nebular gas drag \citep{Rafikov2004,Chambers2008}. The gas drag also causes an orbital migration of the planetesimals. For the gas drag three different regimes are considered: Epstein, Stokes, and quadratic regimes \citep{Rafikov2004,Chambers2008}.

The embryos immersed in the disk grow by accretion of planetesimals in the oligarchic regime and by the accretion of the surrounding gas. For the accretion of planetesimals, we used the prescriptions given by \citet{Inaba2001}. Regarding the accretion of the surrounding gas, if our embryos are able to accrete gas, their gaseous envelopes will grow according to the results obtained by \citet{Guilera2010,Guilera2014}, who solved the classical equations of transport and structure. A similar procedure was implemented in \citet{Miguel2011b}. We also included a limitation on the gas accretion rates considering the capability of the disk to supply mass through the viscous transport \citep{Mordasini2009}. Finally, when the distance between two embryos becomes smaller than 3.5 mutual Hill radii, they merge, considering an inelastic collision. We did not consider type I migration for the planets. \citet{Tanaka2002} found rapid inward type I migration rates in idealized isothermal disks. However, \citet{Alibert2005}, \citet{IdaLin2008}, and \citet{Miguel2011b,Miguel2011a} found that it is necessary to reduce these migration rates using an ad hoc factor to reproduce observations. More recently, \citet{Paardekooper2010,Paardekooper2011} showed that type I migration could substantially change in more realistic disks. Moreover, \citet{BenitezLLambay2015} demonstrated that if the energy released by the planet through accretion of solid material is taken into account, this phenomena generates a heating torque that could significantly slow down, cancel, and even reverse inward type I migration for low-mass planets. This phenomenon occurs in type II
migration when a planet opens a gap in the disk. In our simulations, the only planet that manages to open a gap is the Jupiter analog, but this gap is opened almost at the end of the dissipation of the disk, when there is not enough gas to migrate in a significantly way. Thus, as type II migration does not play an important role in this scenario, we did not consider it. 

Our model of planet formation also includes a radial compositional gradient for the protoplanetary disk. Since the solid surface density presents a discontinuity in the snow line (Eq.~\ref{e3}), we assumed that the two populations of bodies (embryos and planetesimals) located beyond the snow line initially have 50\% of water by mass, while the bodies inside the snow line do not present water. The time evolution of the embryo water contents that is due to the accretion of planetesimals and mergers between them is calculated self-consistently during the evolution of the planetary system. 

\subsection{Initial conditions for $N$-body simulations}

As we mentioned before, the goal of the present work is to analyze the formation and evolution of different planetary systems that host a planet analog to Jupiter, Saturn, Neptune, and a super-Earth just beyond the snow line. To do this, we need to define suitable initial conditions to carry out $N$-body simulations after the gas dissipation. In fact, it is necessary to specify the embryo and planetesimal distributions at the end of the gaseous phase. Although several authors \citep{OBrien2006,Raymond2009,Walsh2011,Ronco2014} used ad hoc initial conditions, \citet{Ronco2015} found that initial conditions obtained from a planet formation model can lead to different and more realistic accretion histories for the planets during the post-oligarchic growth. Thus, we decided to adopt this second mechanism to define the embryo and planetesimal distributions just after the gas dissipation and from this to carry out the $N$-body simulations. 

We here analyzed five different work scenarios that are defined by
\begin{itemize}
\item {\bf } a Jupiter analog that is formed just beyond the snow line at 3 au (scenario I),
\item {\bf } a Saturn analog that is formed just beyond the snow line at 3 au (scenario II), 
\item {\bf } a Neptune analog that is formed just beyond the snow line at 3 au (scenario III), 
\item {\bf } a 5$M_\oplus$ super-Earth that is formed just beyond the snow line at 3 au (scenario IV), and 
\item {\bf } a 2.5$M_\oplus$ super-Earth that is formed just beyond the snow line at 3 au (scenario V).
\end{itemize}

\begin{figure*}[t]
 \centering
 \includegraphics[angle=270, width= 0.95\textwidth]{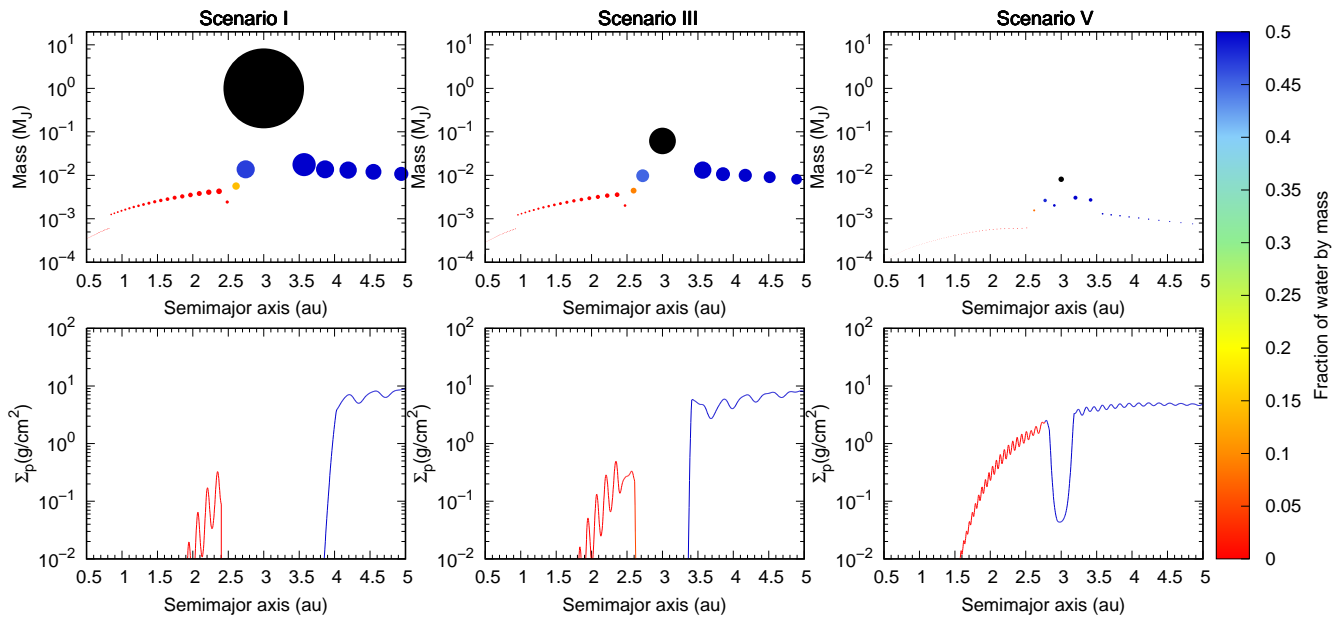}    
 \caption{Embryo distributions (top panels) and surface density of planetesimals (bottom panels) for three of the five scenarios with Jupiter (I), Neptune (III), and super-Earth (V) analogs at the end of the gas phase. The masses of the disks are $\sim 0.11M_\odot$ for the planetary system with a Jupiter analog of $318M_\oplus$, $\sim 0.09M_\odot$ for the planetary system with a Neptune analog of $19.7M_\oplus$ , and $0.04M_\odot$ for a planetary system with a super-Earth analog of $2.5M_\oplus$. The size of the points representing the embryo distributions is scaled by their total mass, except for the most massive planets. The color scale represents the final fraction of water of the embryos with respect to their total masses and of the surface density of planetesimals at the end of the gaseous phase. These embryo and planetesimal distributions were used as initial conditions to develop $N$-body simulations. A color figure is available in the electronic version.}                              
\label{fig:Initial-Conditions}                                                                                                
\end{figure*}

From a series of test numerical simulations developed with the semi-analytical model, we determined the masses of the protoplanetary disks and the suitable photoevaporation rates needed to form each of the five work scenarios defined above. In all cases, we assumed planetesimals of 10 km. Moreover, our simulations considered a viscosity parameter $\alpha= 10^{-4}$ from which the disks dissipate in $\sim$ 5 Myr, which represents a characteristic life timescale for a protoplanetary disk \citep{Mamajek2009,Pfalzner2014}. Our results indicate that the masses of the disks that lead to form scenarios I, II, III, IV, and V are of $\sim 0.11M_\odot$, $\sim 0.10M_\odot$, $\sim 0.09M_\odot$, $\sim 0.05M_\odot$, and $\sim 0.04M_\odot$, respectively. On the one hand, as an example of the formation of the planets in our semi-analytical model, Figure~\ref{fig:Cores-y-Envolturas} shows the time evolution of the masses of the most massive planets of scenarios I (top panel), III (middle panel), and V (bottom panel). The limitation in the gas accretion rate determines the final masses of the Jupiter and Neptune analogs. On the other hand, Figure~\ref{fig:Initial-Conditions} shows the state of the same three scenarios immediately after the dissipation of the gaseous component. These distributions of embryos and planetesimals were used as initial conditions for the $N$-body simulations. We note that we need relative massive disks to form the most massive planets in these systems, especially for scenarios I, II, and III, where Earth-like embryos remain in the inner zone of the disk when the gas dissipates because
we adopted planetesimals of 10~km of radius. As was shown by \citet{Fortier2009,Fortier2013}, the formation of massive cores before the dissipation of the gaseous disk through the accretion of large planetesimals ($r_{\text{p}} \gtrsim 10~$km) in the oligarchic growth regime requires massive disks. This requirement is due to the need to reconcile the formation timescales with the disk
  dissipation timescales. The choice of smaller planetesimals could lead to the formation of massive cores from less-massive disks because smaller planetesimals result in higher planetesimal accretion rates. This could lead to different final configurations of the planetary systems at the end of the gaseous phase and in turn to different long-term evolutions of the systems. However, considering different planetesimal sizes implies the development of a higher number of $N$-body simulations, and because they
have high computational costs, it is beyond the scope of this work.

\section{$N$-body simulations: after the gas phase}

The $N$-body code we used to carry out our study was developed by \citet{Chambers1999} and is known as MERCURY. In particular, we used the hybrid integrator, which uses a second-order mixed variable symplectic algorithm to treat the interaction between objects with separations greater than three Hill radii, and a Burlisch-Stoer method to resolve closer encounters. 

The MERCURY code evolves the orbits of planetary embryos and planetesimals and allows collisions to occur. To reduce the CPU time, our model assumed that embryos interact gravitationally with all other bodies of the simulation, but planetesimals are not self-interacting. Moreover, collisions were treated as inelastic mergers, conserving mass and water content of the interacting bodies.

To use the MERCURY code, it is necessary to specify initial physical and orbital parameters for the most massive planet located just beyond the snow line, planetary embryos, and planetesimals for each of the work scenarios. These initial parameters required for the $N$-body code represent those obtained immediately after the dissipation of the gaseous component, which were derived from the semi-analytical model. 

The initial mass and physical density assumed for the most massive planet of scenarios I, II, III, IV, and V were of 318$M_\oplus$ and 1.33 g cm$^{-3}$, 97.2$M_\oplus$ and 0.7 g cm$^{-3}$, 19.7$M_\oplus$ and 3 g cm$^{-3}$, 5$M_\oplus$ and 3 g cm$^{-3}$, and 2.5$M_\oplus$ and 3 g cm$^{-3}$, respectively. As for the orbital parameters, the most massive planet of each scenario was assigned a circular and planar orbit (i.e. $e=0$ and $i=0$) with an initial semimajor axis of 3 au. Moreover, the initial values of the longitude of the ascending node $\Omega$, the argument of the pericenter $\omega$, and the mean anomaly $M$ were randomly generated between 0$^{\circ}$ and 360$^{\circ}$.

We initially considered 48, 50, 51, 70, and 87 planetary embryos between 0.5 au and 5 au in scenarios I, II, III, IV, and V, respectively. It is worth noting that the initial individual masses of the embryos range between 0.11$M_{\oplus}$ and 5.59$M_{\oplus}$, 0.1$M_{\oplus}$ and 4.88$M_{\oplus}$, 0.09$M_{\oplus}$ and 2.57$M_{\oplus}$, 0.04$M_{\oplus}$ and 1.65$M_{\oplus}$, and 0.03$M_{\oplus}$ ,
and 0.97$M_{\oplus}$ for scenarios I, II, III, IV, and V, respectively. Moreover, for any case, we assumed physical densities of 3 g cm$^{-3}$ for all planetary embryos. In particular, the top panels of Figure~\ref{fig:Initial-Conditions} show the initial masses of planetary embryos as a function of the initial semimajor axis for scenarios I, III, and V, which were derived from the semi-analytical model. Regarding the orbital parameters, we randomly generated initial eccentricities and inclinations lower than 0.02 and 0.5$^{\circ}$, respectively, for all planetary embryos. Moreover, the initial values of the longitude of the ascending node $\Omega$, the argument of the pericenter $\omega$, and the mean anomaly $M$ were randomly generated between 0$^{\circ}$ and 360$^{\circ}$.   

For each of the five scenarios under consideration, we included 500 planetesimals, each of which was assigned a physical density of 1.5 g cm$^{-3}$. The initial individual mass of each planetesimal was of 0.0146$M_\oplus$, 0.0176$M_\oplus$, 0.0194$M_\oplus$, 0.0184$M_\oplus$, and 0.0172$M_\oplus$ for scenarios I, II, III, IV, and V, respectively. For the orbital parameters, the initial semimajor axes of the planetesimals were generated using the acceptance-rejection method developed by John von Neumann. To do this, we used the planetesimal surface density profiles at the end of the gaseous phase for each scenario. In particular, the bottom panels of Figure~\ref{fig:Initial-Conditions} show such planetesimal surface density profiles as a function of the initial semimajor axis for scenarios I, III, and V, which were derived from the semi-analytical model. Just like for embryos, the planetesimals were randomly assigned initial eccentricities and inclinations lower than 0.02 and 0.5$^{\circ}$, respectively. Moreover, the initial values of the longitude of ascending node $\Omega$, the argument of pericenter $\omega$, and the mean anomaly $M$ were randomly chosen between 0$^{\circ}$ and 360$^{\circ}$.

We recall that the main goal of our research is to analyze the formation and evolution of potentially habitable planets. In general terms, we considered that the most important condition required for a planet to be habitable is the permanent presence of liquid water on its surface. \citet{Kopparapu2013b} defined conservative inner and outer edges for the HZ, which are determined by loss of water and by the maximum greenhouse effect provided by a $\text{CO}_{2}$ atmosphere, respectively. For a Sun-like star, they computed a conservative estimate width of the HZ of 0.99-1.69 au. The authors also determined optimistic inner and outer limits for the HZ. For the inner limit, an optimistic estimate is based on the inference that Venus has not had any liquid water on its surface for at least the past one billion years \citep{SolomonHead1991}. For the outer limit, an optimistic empirical limit is estimated based on the observation that early-Mars was warm enough for liquid water to flow on its surface \citep{Pollack1987,Bibring2006}. For a Sun-like star, \citet{Kopparapu2013b} computed an optimistic estimate for the width of the HZ of 0.75-1.77 au. From these estimates, we assumed that a planet is in the HZ of the system and so could maintain liquid water on its surface if its whole orbit is contained inside the optimistic edges, that is, if it has a perihelion distance $q \geq$ 0.75 au and an aphelion distance $Q \leq$ 1.77 au.

However, we considered that it seems too conservative to require that the perihelion and aphelion distances are both inside the HZ for a planet to be habitable. \citet{Williams2002} showed that provided that an ocean is present to act as a heat capacitor,  primarily the time-averaged flux affects the habitability over an eccentric orbit. Planets with high orbital eccentricities ($e >$ 0.1) have a higher average orbital flux, which may help eccentric planets near the edges of the HZ to maintain habitable conditions. We took this criterion into account for eccentric planets located near the edges of the HZ with orbits that are not fully contained in that region.

Owing to the stochastic nature of the accretion process, we carried out 15 $N$-body simulations for each of scenarios I and II, 10 $N$-body simulations for scenario III, and 6 $N$-body simulations for each of scenarios IV and V. We integrated each simulation for 200 Myr, which is a good choice as an upper limit for the formation timescale of the terrestrial planets of our solar system \citep{Jacobson2014}. Moreover, we used a time step of six days to carry out the integrations, which is shorter than 1/20th of the orbital period of the innermost body in the simulation. Finally, to avoid any numerical error for small-perihelion orbits, a non-realistic size for the solar radius of 0.1 au was adopted in each $N$-body simulation.

\section{Results}

Here, we present results of $N$-body simulations concerning the formation and evolution of terrestrial-like planets in the HZ and water delivery around Sun-like stars in different dynamical environments. We wish to study these processes in systems harboring different planets that act as main gravitational perturbers around the snow line. Based on this, we defined five different scenarios
that have a planet analog to Jupiter (I), Saturn (II), Neptune (III), and a super-Earth with 5$M_{\oplus}$ (IV) and 2.5$M_{\oplus}$ (V) at 3 au. In the following, we present the main dynamical and physical properties of the terrestrial-like planets that
survive in the HZ for each of our scenarios.

\subsection{Scenario I: Jupiter}

In this first dynamical scenario, a giant Jupiter-mass planet\footnote{From now on, we refer to the Jupiter-mass planet as Jupiter.} of 318$M_{\oplus}$ was formed at 3 au during the gaseous phase. Because of the stochastic nature of the accretion process, we performed 15 $N$-body simulations, 10 of which have formed one planet in the HZ after 200 Myr of evolution. Table \ref{tab:JPlanets} lists the main physical properties and orbital parameters of each planet of interest\footnote{We say that a planet is a ``planet of interest'' if it survived in the HZ at the end of the simulation.} formed in this scenario. 

\begin{table*}
\caption{Physical and orbital properties of the planets of interest formed in scenario I. $a_{\text{i}}$ and $a_{\text{f}}$ are the initial and final semimajor axes in au, respectively, $M_{\text{i}}$ and $M_{\text{f}}$ the initial and final masses in $M_{\oplus}$, respectively, $T_{\text{GI}}$ is the timescale in Myr associated with the last giant impact produced by an embryo, and $W$ is the final percentage of water by mass after 200 Myr of evolution. The planets of interest in SIMs 1, 3, 5, and 9 did not receive any impact during the simulations.
}
\begin{center}
\begin{tabular}{|c|c|c|c|c|c|c|c|}
\hline
\hline
Simulation & $a_{\text{i}}$(au)  & $a_{\text{f}}$(au) & $M_{\text{i}}(M_{\oplus})$ & $M_{\text{f}}(M_{\oplus})$ & $T_{\text{GI}}$ (Myr) & $W(\%)$ \\
\hline
\hline
SIM 1  & 3.57 & 1.51 & 5.63 & 5.63 & --  & 50 \\
\hline
SIM 2  & 2.74 & 1.02 & 4.37 & 6.35 & 4.0 & 34 \\
\hline
SIM 3  & 3.57 & 1.18 & 5.63 & 5.63 & --  & 50 \\
\hline
SIM 4  & 3.57 & 0.90 & 5.63 & 6.44 & 2.6 & 43 \\
\hline
SIM 5  & 3.57 & 1.07 & 5.63 & 5.63 & --  & 50 \\
\hline
SIM 6  & 3.57 & 1.56 & 5.63 & 6.81 & 1.3 & 41 \\
\hline
SIM 7  & 4.54 & 0.89 & 3.83 & 7.53 & 3.4 & 25 \\
\hline
SIM 8  & 1.38 & 0.81 & 0.72 & 4.53 & 3.8 & 0 \\
\hline
SIM 9  & 1.16 & 1.68 & 0.59 & 0.59 & --  & 0 \\
\hline
SIM 10 & 1.54 & 0.91 & 0.84 & 5.0 & 2.9  & 0 \\
\hline

\end{tabular}
\end{center}
\label{tab:JPlanets}
\end{table*}

\begin{figure*}[ht]
 \centering
 \includegraphics[angle=0, width= 0.45\textwidth]{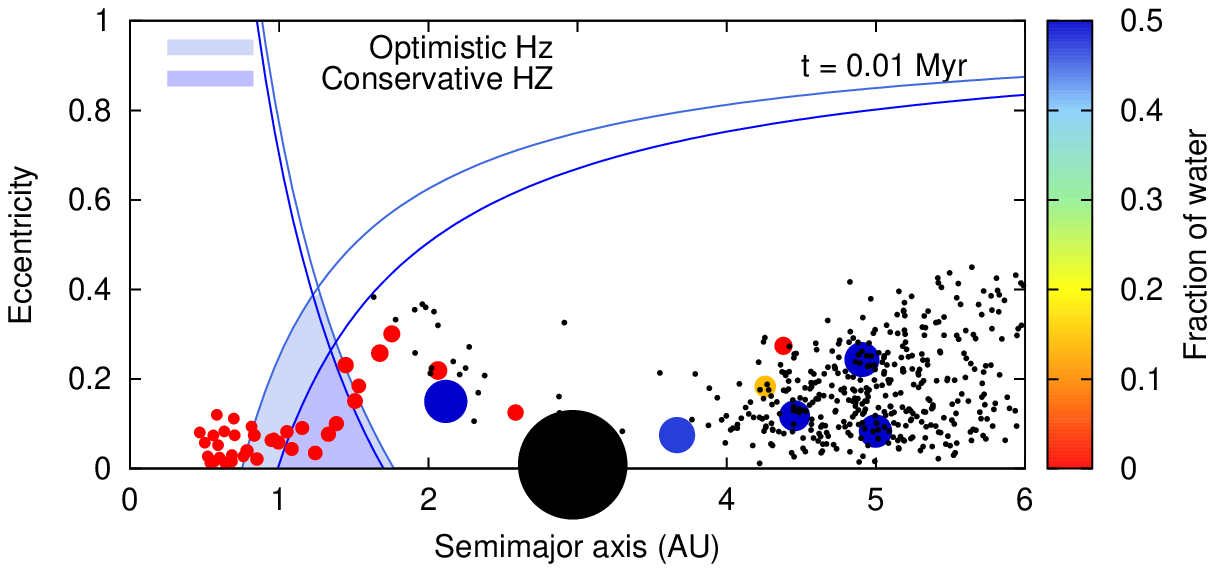}   
 \includegraphics[angle=0, width= 0.45\textwidth]{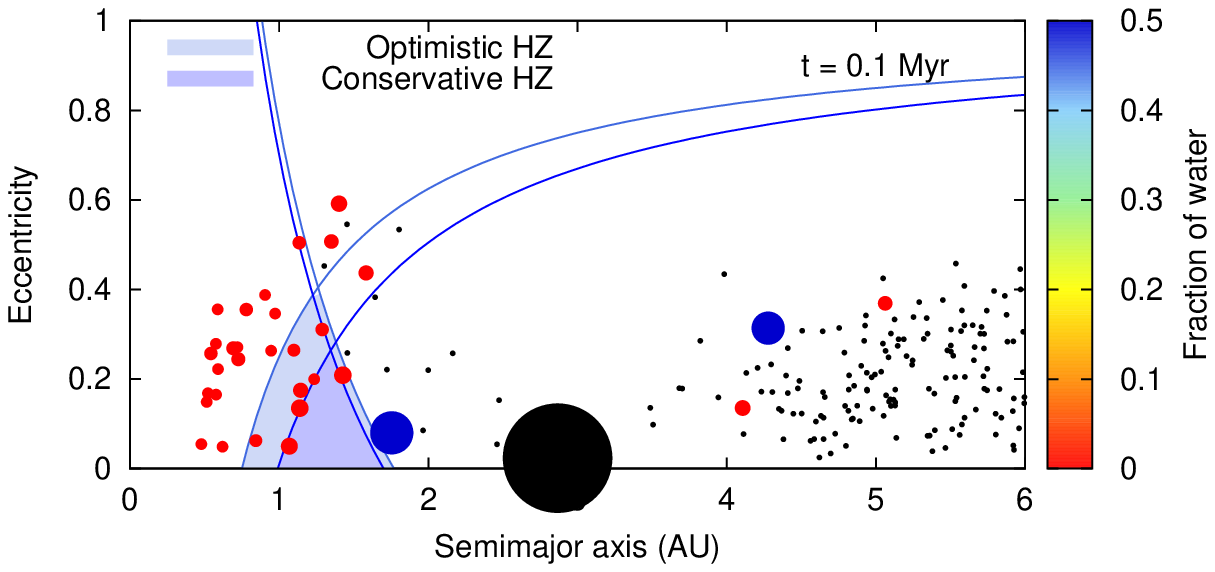}\\  
 \includegraphics[angle=0, width= 0.45\textwidth]{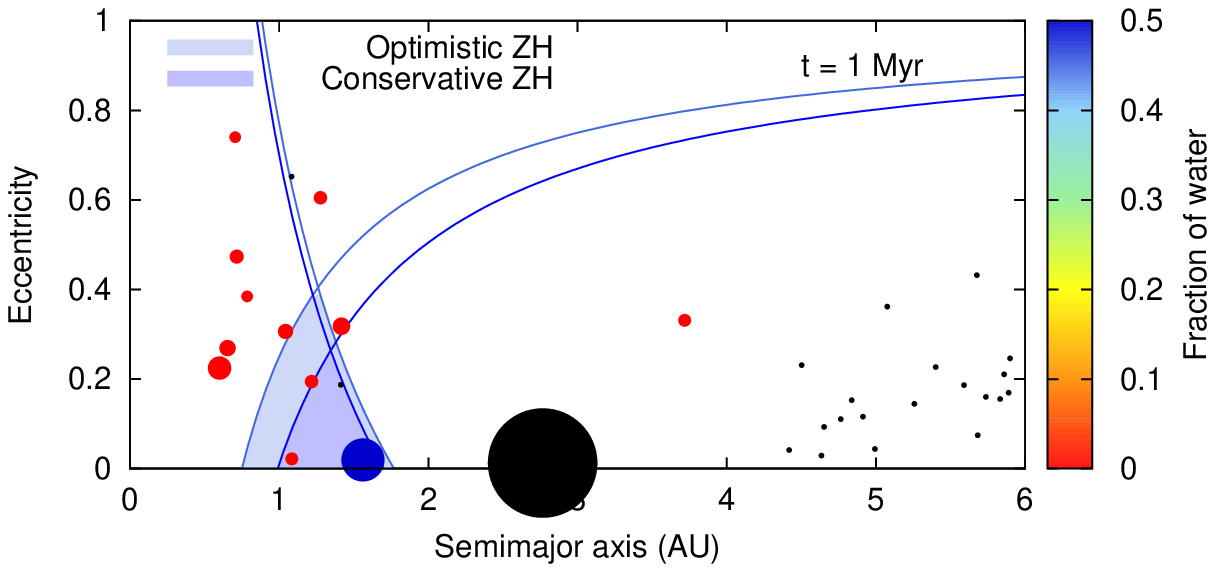}   
 \includegraphics[angle=0, width= 0.45\textwidth]{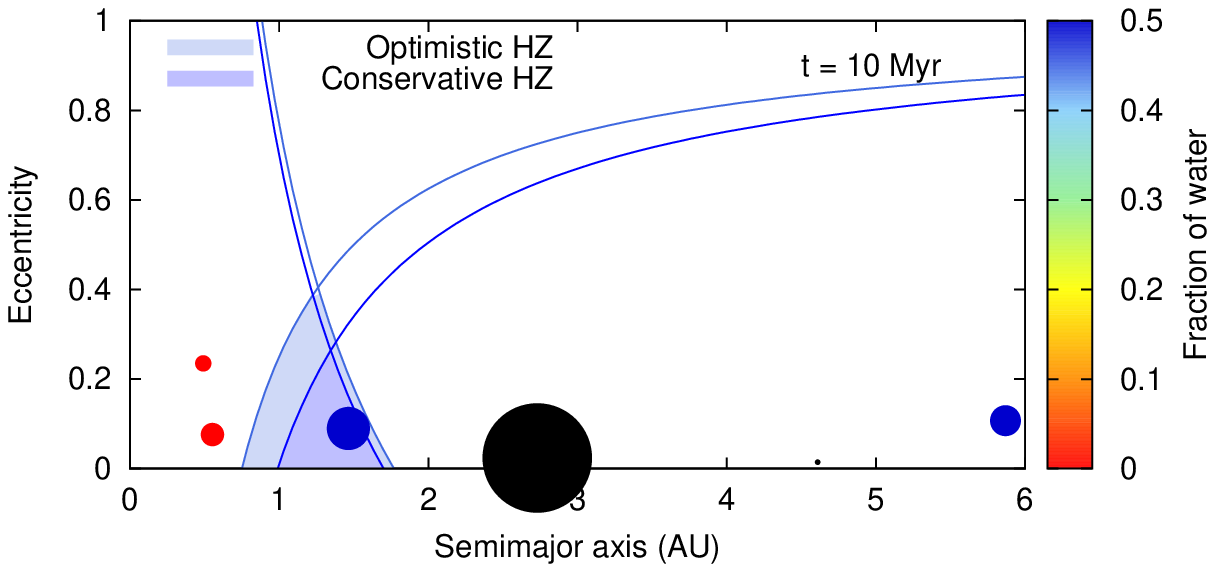}\\ 
 \includegraphics[angle=0, width= 0.45\textwidth]{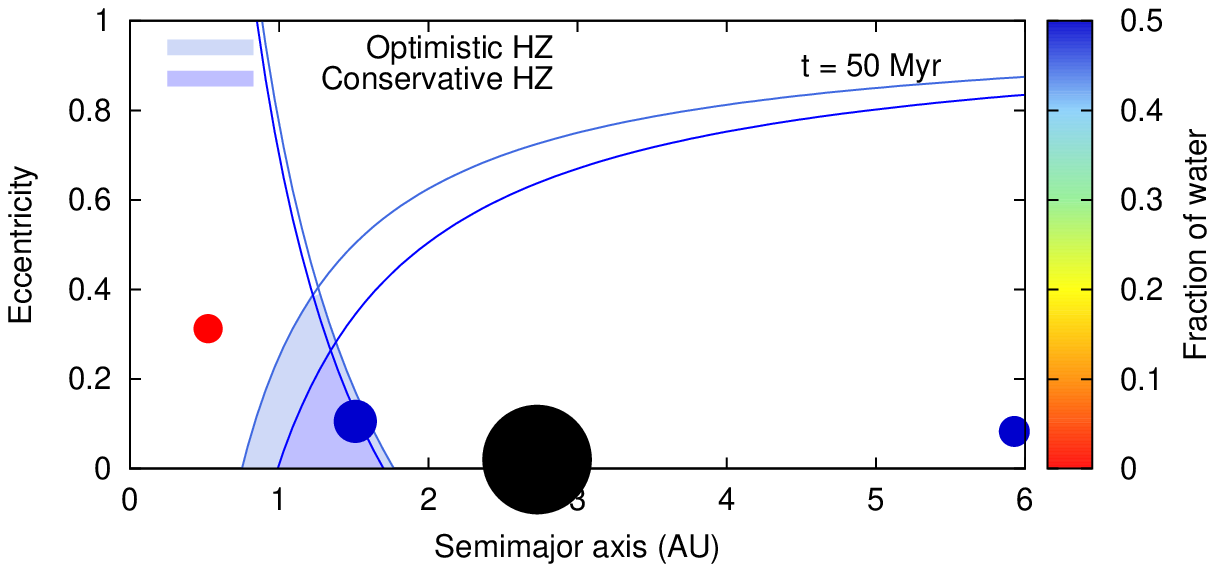}   
 \includegraphics[angle=0, width= 0.45\textwidth]{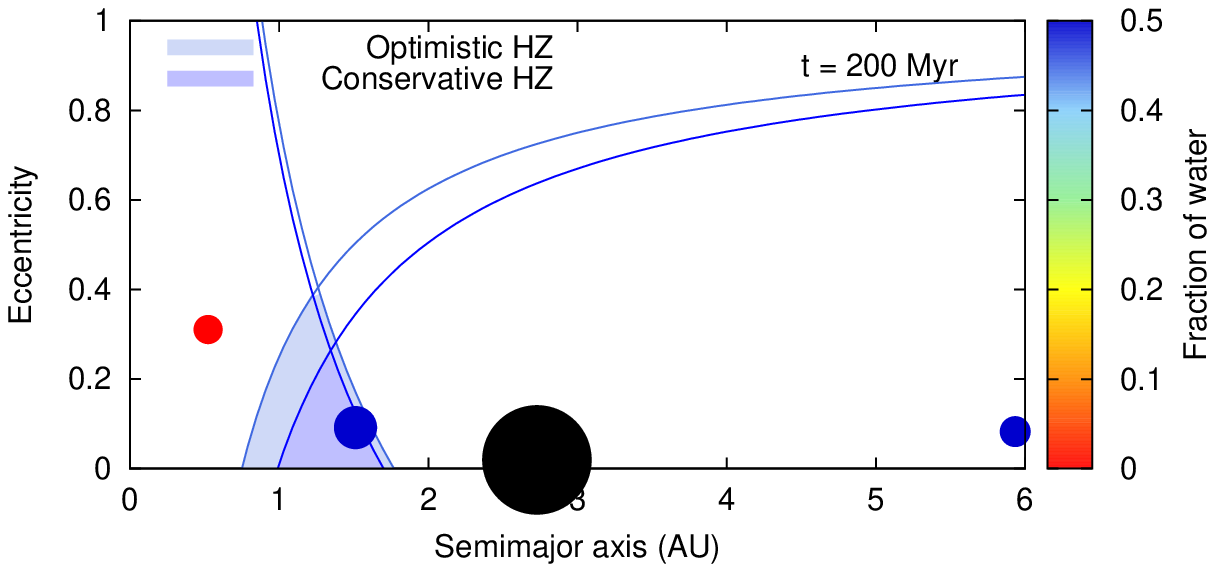}    
 \caption{Evolution in time of SIM 1, corresponding to scenario I. Planetary embryos are plotted as colored circles following the color scale that represents the fraction of water with respect to their total masses. The blue and light blue shaded areas represent the conservative and the optimistic HZ respectively. Moreover, the blue and light blue curves represent curves of constant perihelion and aphelion for the conservative and the optimistic HZ. At the end of the simulation, a super-Earth of 5.63 $M_{\oplus}$ and 50\% water by mass survives in the HZ. A color version of this figure is only available in the electronic version.}
\label{fig:JSnapshot}
\end{figure*}

{\bf General evolution of the system:} The dynamical evolution of the simulations that produce planets in the HZ is very similar. Thus, we analyze the results of one of them as representative of the whole group. We propose SIM 1. Figure~\ref{fig:JSnapshot} shows six snapshots in time on the semimajor axis-eccentricity plane of a system featuring Jupiter as the most massive planet around the snow line.

\begin{figure}[ht]
 \centering                    
 \includegraphics[angle=0, width= 0.45\textwidth]{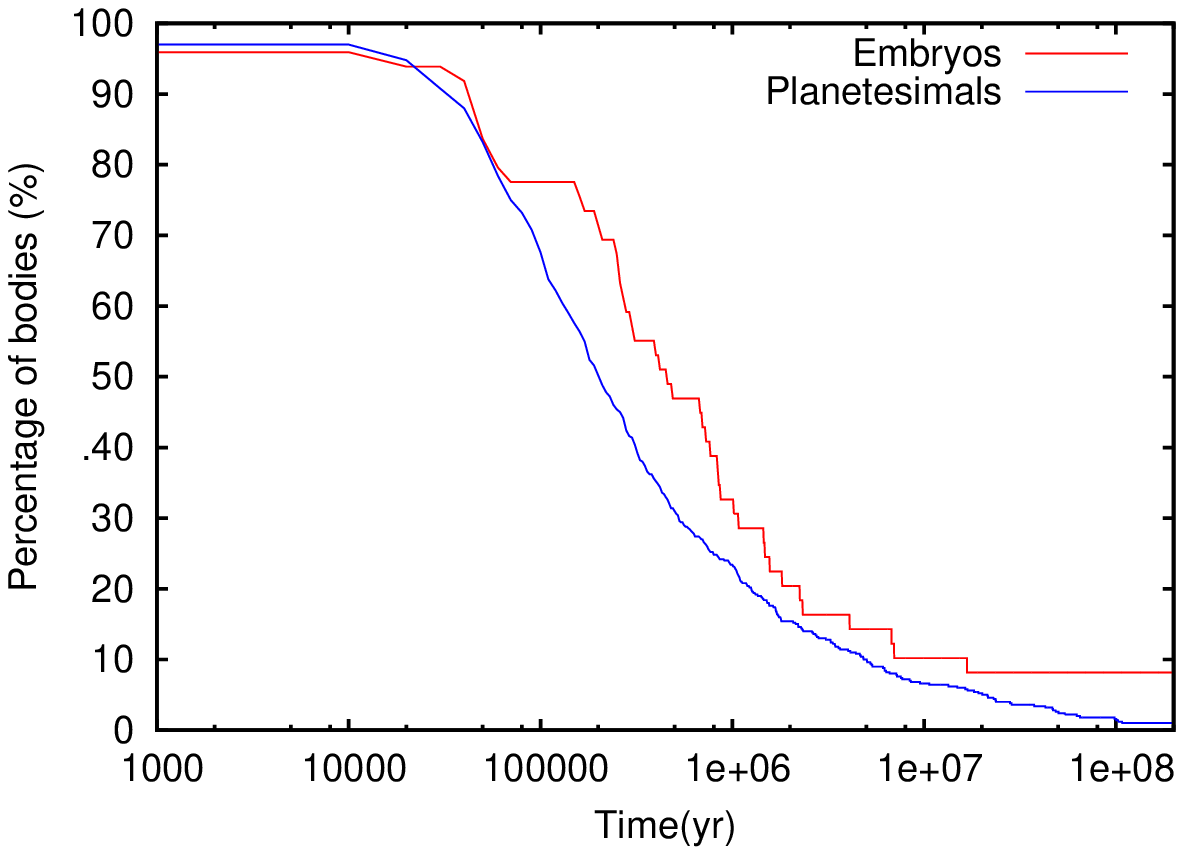}\\
 \includegraphics[angle=0, width= 0.45\textwidth]{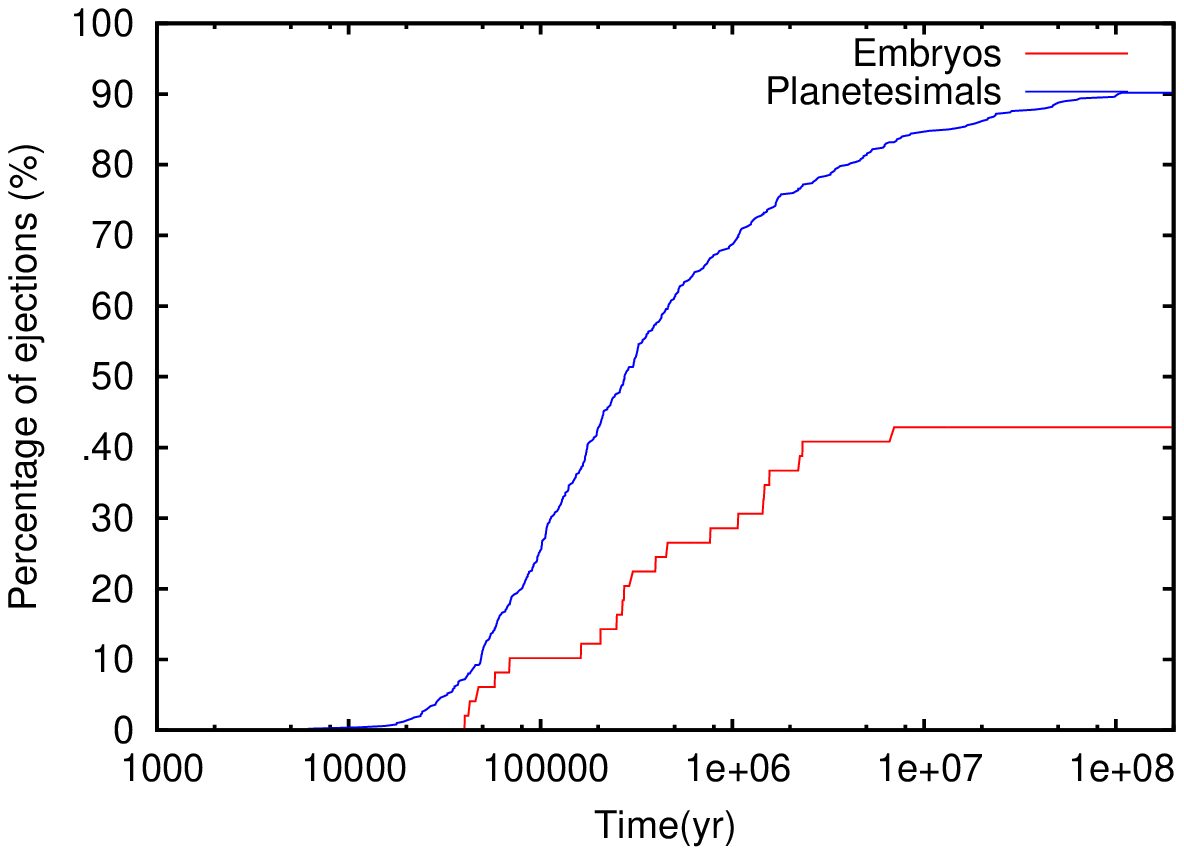} 
 \caption{Percentage of planetary embryos (red curve) and planetesimals (blue curve) that survive in the system (top panel) and are ejected (bottom panel) from the system as a function of time for scenario I.}
\label{fig:NTJup}     
\end{figure}

In general terms, the overall progression of this simulation can be described as follows. From the beginning, the planetary embryos and planetesimals were quickly excited by Jupiter's gravitational perturbations. The eccentricities of embryos and planetesimals therefore increased until their orbits crossed and accretion collisions occurred. Thus, planetary embryos grew by accretion of other embryos and planetesimals. However, a very large fraction of embryos and planetesimals were removed from the system on very early timescales. The two panels of Figure~\ref{fig:NTJup} show that 70\% of the planetesimals were ejected from the system in 1 Myr, while only 30\% of the embryos survived in the system by that time. 

By the end of the simulation, less than 10\% of the planetary embryos survived in the system, while the planetesimals were completely removed. Finally, after 200 Myr of evolution, a super-Earth of $\sim$ 5.63$M_{\oplus}$ survived in the HZ with 50\% of water by mass, while Jupiter remained  at around the snow line. 

{\bf Orbital evolution}: Table~{\ref{tab:JPlanets}} shows the initial semimajor axis of the accretion seeds of the planets surviving in the HZ after 200 Myr of evolution\footnote{Following \citet{Raymond2009}, we define a planet accretion seed as the largest embryo involved in its collisional history.}. According to this, these planets have two different origins depending on whether their accretion seeds were initially located outside or inside the snow line at 2.7 au.

\begin{figure}[ht]
 \centering                    
 \includegraphics[angle=0, width= 0.45\textwidth]{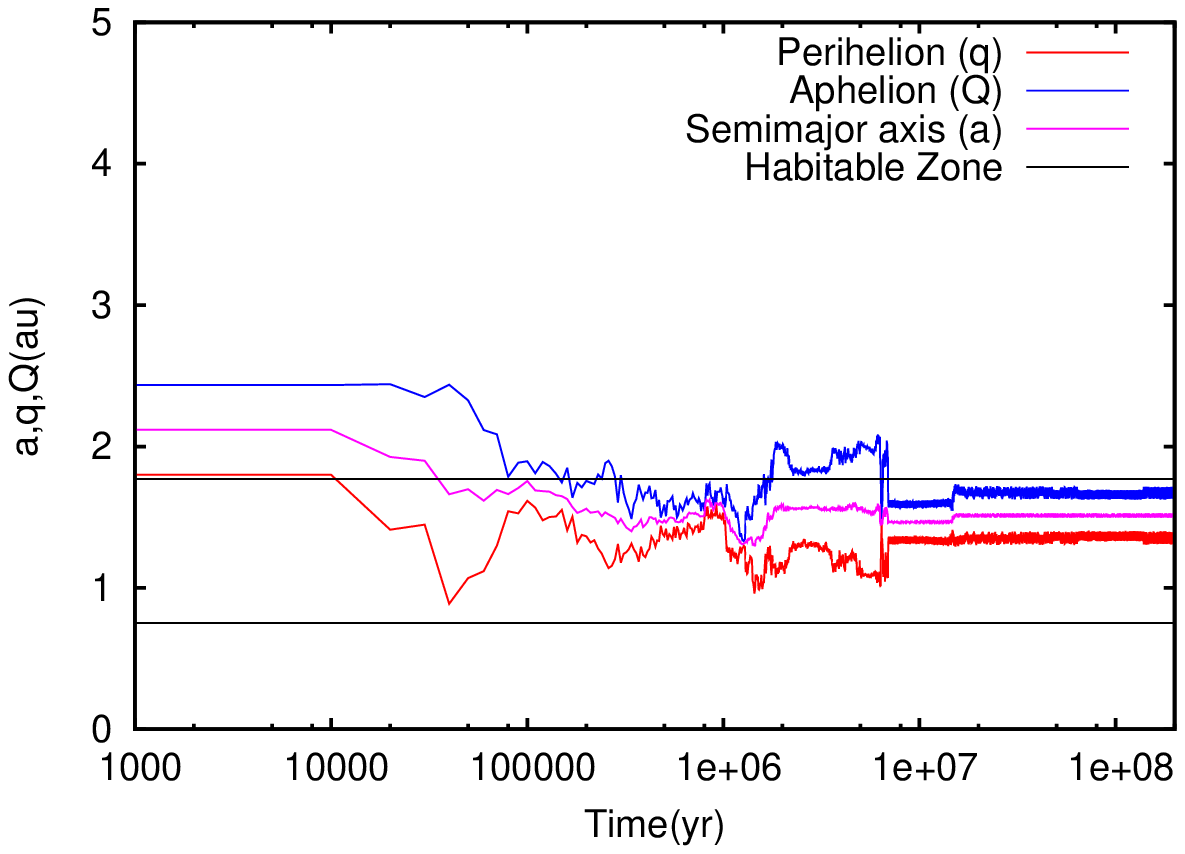}\\
 \caption{Evolution in time of the semimajor axis $a$, the perihelion $q,$ and aphelion $Q$ distances for the planet that survives in the HZ in SIM 1, corresponding to scenario I. The black lines indicate the limits of the optimistic HZ. We remark that the planet started the simulation located at 3.57 au and was abruptly scattered inward in 1000 yr. A color version of this figure is available in the electronic version of the journal.}
\label{fig:OrbitJup}
\end{figure}

On one hand, we find that seven planets (SIMs 1-7) evolved from accretion seeds that were located beyond the snow line at the end of the gaseous phase.
In particular, Figure~\ref{fig:OrbitJup} shows the evolution in time of the semimajor axis, and the perihelion and aphelion distances for the planet of interest resulting from SIM 1. This planet started its evolution beyond the snow line, at 3.57 au, and was abruptly scattered inward in less than 1000 yr after crossing the orbit of Jupiter. Then, the planet semimajor axis reached the HZ in $\sim$50000 yr and survived with its orbit fully contained inside the limits of the HZ after $\sim$8 Myr. The planets that survived in the HZ in SIMs 1-7 shown in Table~{\ref{tab:JPlanets}} evolved following similar behaviors. The orbits of all but one of the planets are fully contained within the limits of the HZ. However, the planet of interest in SIM 6 reached aphelion distances greater than the outer limit of the HZ with a semimajor axis of 1.56 au and an eccentricity of 0.18. We consider this planet a planet of interest following the average flux criterion given by \citet{Williams2002}.

On the other hand, three planets (SIMs 8-10) evolved from accretion seeds that were located inside the snow line at the end of the gas phase. 
In particular, we find that their orbits were confined in the HZ region during the 200 Myr of evolution.

{\bf Mass evolution}: Table~\ref{tab:JPlanets} shows the initial and final masses of the ten planets of interest that formed in this scenario. Their primordial masses range between 3.83$M_{\oplus}$ and 5.63$M_{\oplus}$ for the planets that evolved from accretion seeds from outside the snow line, and between 0.59$M_{\oplus}$ and 0.84$M_{\oplus}$ for those with accretion seeds located inside the snow line. After 200 Myr of evolution, these planets are super-Earths with masses ranging between 4.53$M_{\oplus}$ and 7.53$M_{\oplus}$, and one of them is a sub-Earth of 0.59$M_{\oplus}$. It is very interesting to analyze how these masses were obtained for each of these planets of interest.

\begin{figure}[ht]
 \centering
 \includegraphics[angle=0, width= 0.45\textwidth]{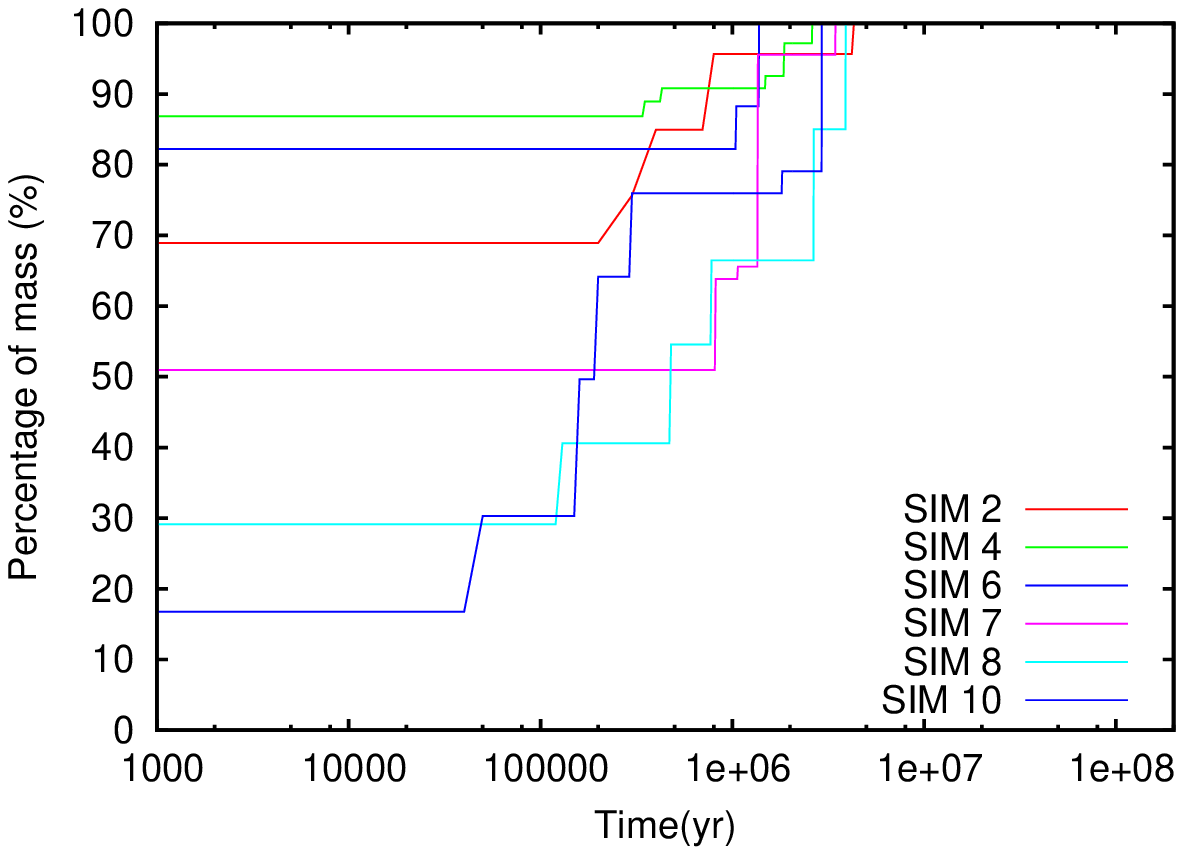}\\
 \includegraphics[angle=0, width= 0.45\textwidth]{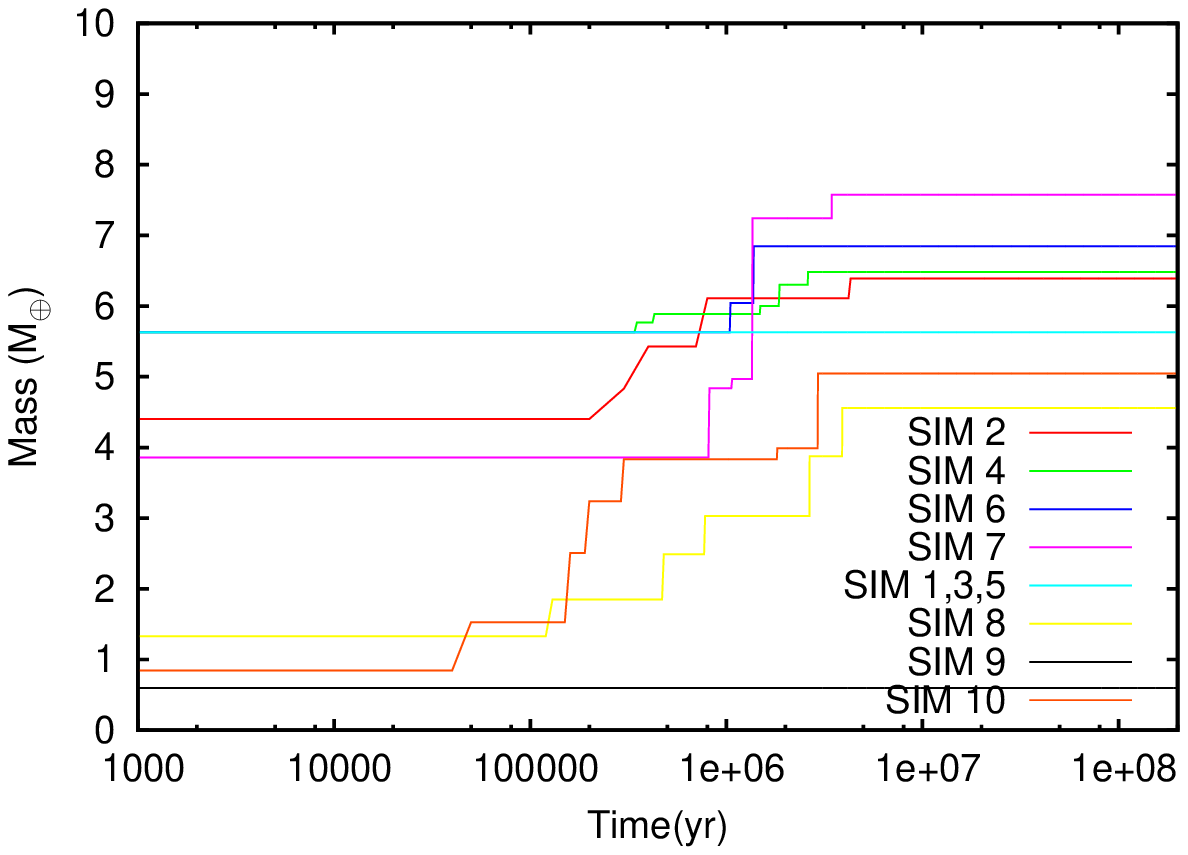}
 \caption{Percentage of mass (top) and total mass (bottom) as a function of time of the planets of interest in scenario I. A color version of this figure is available in the electronic version of the journal.}
 \label{fig:MassJup}
\end{figure}

Figure~\ref{fig:MassJup} shows the evolution in time of the mass (bottom panel) and percentage of mass (top panel) of the planets surviving in the HZ. We find that each of these planets received very few impacts along their evolution; the maximum number of impacts received for a given planet is six. Moreover, it is important to remark that they were only hit by planetary embryos. Collisions with planetesimals are not registered. The top panel of Figure~\ref{fig:MassJup} shows that in general terms, at least 70\% of the final masses are primordial for the planets that started their evolution outside the snow line. Moreover, the planets of interest resulting from SIMs 1, 3, 5, and 9 did not receive any impact during their evolution, which means that their masses are entirely primordial. However, the planets formed in the HZ from SIMs 8 and 10, which started their evolution inside the snow line, accreted more than $\sim$80\% of their final masses by impacts with embryos.  

The planets surviving in the HZ have formation timescales of $\sim$4 Myr, which is defined as the time when they received the last impact by a planetary embryo. This result is consistent with the fact that most of embryos and planetesimals were removed from the system in $\sim$1 Myr. We also wish to compare the formation timescales of the planets of interest produced in the present scenario with that associated to the formation of our home planet Earth. According to the generally accepted scenario, the last giant impact on Earth formed the Moon and initiated the final phase of core formation by melting Earth's mantle. Using highly siderophile element abundance measurements, \citet{Jacobson2014} determined a Moon-formation age of 95$\pm$32 Myr after the condensation of the first solids in the solar system. Clearly, the planets of interest formation timescales in this scenario are very short compared to the timescale associated to Earth. 

{\bf Water delivery}: The last physical feature that we wish to study is the water content of all the planets that formed in the HZ in this dynamical scenario. In our research, the water content of a planet has two possible sources: primordial (accreted during the gas phase), and collisional (accreted by impacts during the post-gas phase). In order to fulfill this task, we wish to determine the primordial water content and identify the nature of the bodies that collided with the planets during the 200 Myr evolution after the dissipation of the gaseous component. These bodies can be planetesimals, dry planetary embryos (formed within the snow line), and water-rich planetary embryos (formed beyond the snow line).

We described above that all the impactors on the planets surviving in the HZ in the present scenario were planetary embryos. Moreover, it is worth remarking that all the impactors were dry planetary embryos. Thus, the planets that survived in the HZ did not accrete any water during their collisional evolution after the gas dissipation. Then, if these planets have any water content, it is entirely primordial. This means that the planets of interest that originated outside the snow line have very high final water contents (more than 25\% of water by mass), while the planets that were formed in situ in the HZ are dry planets. \\

According to their physical properties and orbital parameters, the planets that survived in the HZ in scenario I can be classified into two categories.
\begin{itemize}
\item {\it Water worlds:} they are represented by the planets surviving in the HZ in SIMs 1-7. Such planets started their evolution in a water-rich region beyond the snow line and ended with very high water contents by mass. It is worth remarking that these final water contents are entirely primordial.
\item {\it Planets in situ:} they are represented by the planets formed in the HZ in SIMs 8-10. These planets started their evolution in a dry region inside the snow line. Since these planets did not accrete any water during the evolution, they ended as dry worlds. 
\end{itemize}

The results analyzed in this section allow us to suggest that the formation of planets in the HZ is a very efficient process in scenario I:
10 of 15 $N$-body simulations produced planets in the HZ, 7 of which contained water after 200 Myr of evolution. However, it is worth remarking that this scenario was not able to form planets with physical and dynamical properties similar to the Earth. 

\subsection{Scenario II: Saturn}

Here, we present the results of numerical simulations in which a Saturn-like planet\footnote{From now on, we refer to this Saturn-mass planet as Saturn.} with a mass of 97.2$M_{\oplus}$ was formed at 3 au at the end of the gaseous phase. Owing to the stochastic nature of the accretion process, we performed 15 $N$-body simulations, 4 of which formed one planet in the HZ. The physical and orbital properties of the planets of interest formed in this scenario are listed in Table~\ref{tab:SatPlanets}. We find that the general evolution of the system in these four simulations is very similar to those that formed water worlds in scenario I. 

\begin{table*}
\caption{Physical and orbital properties of the planets of interest formed in scenario II. $a_{\text{i}}$ and $a_{\text{f}}$ are the initial and final semimajor axes in au, respectively, $M_{\text{i}}$ and $M_{\text{f}}$ the initial and final masses in $M_{\oplus}$, respectively, $T_{\text{GI}}$ is the timescale in Myr associated with the last giant impact produced by an embryo, and $W$ is the final percentage of water by mass after 200 Myr of evolution.}
\begin{center}
\begin{tabular}{|c|c|c|c|c|c|c|c|}
\hline
\hline
Simulation & $a_{\text{i}}$(au)  & $a_{\text{f}}$(au) & $M_{\text{i}}(M_{\oplus})$ & $M_{\text{f}}(M_{\oplus})$ & $T_{\text{GI}}$ (Myr) & $W(\%)$ \\
\hline
\hline
SIM 1  & 4.17 & 1.03 & 3.67 & 5.34 & 3.1 & 35 \\
\hline
SIM 2  & 3.86 & 1.12 & 3.86 & 7.05 & 1.3 & 48 \\
\hline
SIM 3  & 3.86 & 1.33 & 3.86 & 4.04 & 0.3 & 48 \\
\hline
SIM 4  & 4.17 & 1.08 & 3.67 & 5.40 & 3.4 & 34 \\
\hline\end{tabular}
\end{center}
\label{tab:SatPlanets}
\end{table*}

\begin{figure}[ht]         
 \centering                    
 \includegraphics[angle=0, width= 0.45\textwidth]{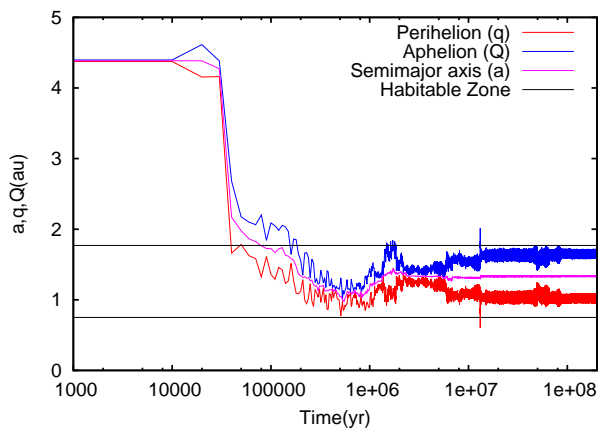}\\
 \caption{Evolution in time of the semimajor axis, perihelion, and aphelion distances for the planet that survived in the HZ in SIM 1, corresponding to scenario II.}
\label{fig:OrbitSat}
\end{figure}

{\bf Orbital evolution:} In the beginning of the simulations, the accretion seeds of the planets of interest were located beyond the snow line at 3.86 au (SIMs 2 and 3) and 4.17 au (SIMs 1 and 4). In particular, Figure~\ref{fig:OrbitSat} shows the evolution in time of the semimajor axis, perihelion, and aphelion distances of the planet of interest formed in SIM 3, as representative of the whole group of simulations. In general terms, the planets migrated into the HZ very abruptly in $\sim$0.1 Myr as a result of interactions with other embryos. Afterward, the orbit stabilized inside the HZ at $\sim$10 Myr and remained in that region until the end of the simulation. It is worth mentioning that the planets of interest resulting from SIMs 2 and 4 had slightly smaller perihelion distances than the inner limit of the optimistic HZ during the late stages of evolution. In spite of this, since the planets of SIMs 2-4 have maximum eccentricities of 0.33 and 0.54, respectively, we still consider them as planets of interest following the criterion given by \citet{Williams2002}.  

\begin{figure}[ht]            
  \centering
 \includegraphics[angle=0, width= 0.45\textwidth]{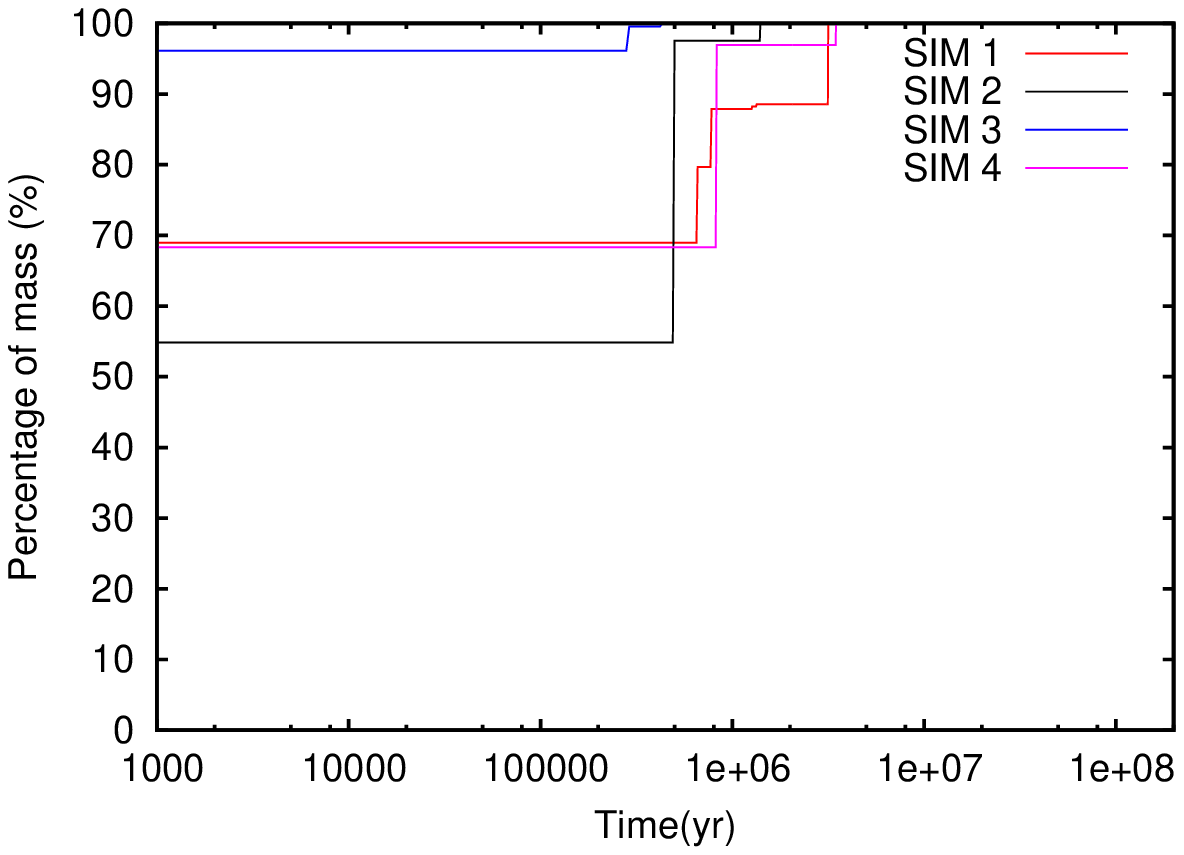}\\
 \includegraphics[angle=0, width= 0.45\textwidth]{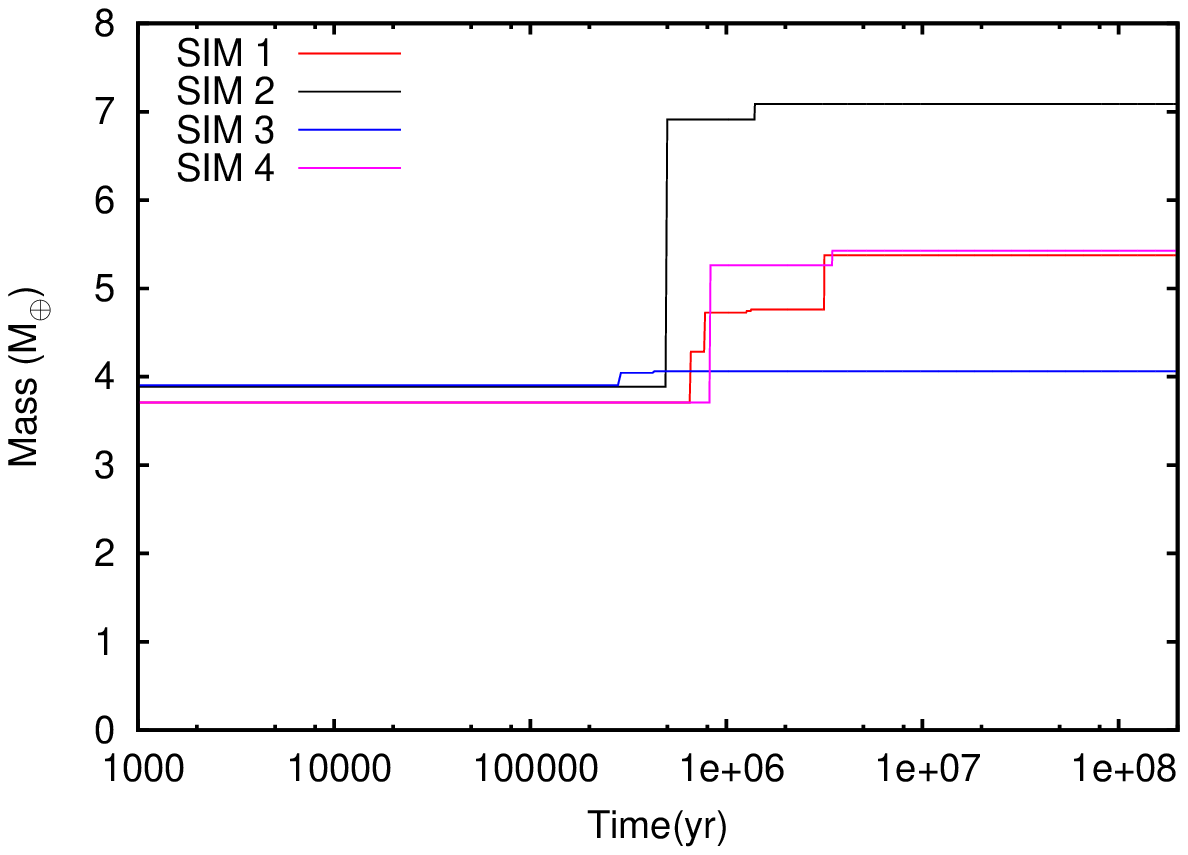}
 \caption{Percentage of mass (top) and total mass (bottom) as a function of time of the planets of interest in scenario II. A color version of this figure is available in the electronic version of the journal.}  
\label{fig:MassSat}
\end{figure}

{\bf Mass evolution:} The four planets of interest that formed in this scenario are super-Earths with final masses between 4.04$M_{\oplus}$ and 7.05$M_{\oplus}$. We are interested in determining how these masses were obtained in time. Figure~\ref{fig:MassSat} shows the evolution in time of the mass (bottom panel) and the percentage of mass (top panel) of the planets that survived in the HZ. Table~{\ref{tab:SatPlanets}}
shows that the planets of interest evolved from two accretion seeds with primordial masses of 3.67$M_{\oplus}$ (SIMs 1 and 4) and 3.86$M_{\oplus}$ (SIMs 2 and 3). Just like in scenario I, each planet that survived in the HZ received a small number of impacts; the maximum number of total impacts is six. The impactors were mostly planetary embryos, but the planets were also hit by planetesimals. However, since each planet only collided with up to three planetesimals, the total mass contribution from planetesimals is too low to be considered relevant in the final planetary mass. In this scenario, at least half of the final planetary mass is primordial. Figure~\ref{fig:MassSat} shows that the planet that
formed in SIM 2 received 45\% of its final mass in a giant impact with a planetary embryo, while the planets formed in the remaining three simulations have primordial mass percentages of at least $\sim$70\%.

Our numerical simulations also show that all the impacts with embryos received by the planets that survived in the HZ occurred within the first 4 Myr of evolution, just like in scenario I. Thus, the planets of interest have very short formation timescales in comparison with the timescale associated to the formation of Earth.

{\bf Water delivery:} The four planets that survived in the HZ in this scenario evolved from accretion seeds that were initially located outside the snow line. This means that their primordial water content is 50\% of their initial mass. Most of the impactors on these planets were inner planetary embryos. However, during their evolution, they were also hit by some water-rich bodies, both planetary embryos and planetesimals. 

The number of impacts with planetesimals is smaller than three in all simulations, meaning that the planetesimal contribution to the water content is also irrelevant compared to the primordial water. Only the planet resulting from SIM 2 received a giant impact by a water-rich planetary embryo. The water contribution of an impact like this is very important because it is about half of the planet's final water content. However, if the planets were only hit by planetesimals and not water-rich embryos, which is the case for the planets that formed in SIMs 1, 3 and 4, then 99\% of their final water content would be primordial. According their physical and dynamical properties, the four planets that survived in the HZ in this scenario are water worlds with masses between 4.04$M_{\oplus}$ and 7.05$M_{\oplus}$ and final water contents of 34\%-48\% by mass. \\

These results suggest that this dynamical scenario is not as efficient as scenario I for the formation of planets in the HZ. Four out of 15 $N$-body simulations produced planets in the HZ, each of which is a water world. Like scenario I, the systems harboring a Saturn-like planet as main perturber were not able to form planets with physical and dynamical properties similar to those of the Earth.

Comparing these first two systems under consideration, scenario I seems to be more efficient to form planets in the HZ than scenario II: 10 out of 15 planets were formed in the HZ in scenario I, while only 4 out of 15 formed in scenario II. However, we find that this result depends on the definition used for the HZ. When we use the conservative limits defined by \citet{Kopparapu2013b} instead of the optimistic limits that we used here, we find that both scenarios are equally efficient. Five planets of interest were formed in the conservative HZ in scenario I, while four were produced in scenario II. The role of giant planets in the dynamical evolution of terrestrial planets is a topic of special interest. We will continue this research in a future work. 

\subsection{Scenario III: Neptune}

In this section, we present the numerical results of the third dynamical scenario, in which a Neptune-like planet\footnote{From now on, we refer to this planet as Neptune.} of 19.7$M_{\oplus}$ is formed at 3 au at the end of the gaseous phase and acts as main gravitational perturber of the system. We performed 10 simulations, 7 of which  have formed one planet in the HZ. Table~\ref{tab:NepPlanets} lists the main physical properties and orbital parameters of these planets. We clearly see that the planet that survived in the HZ in all 7 simulations is Neptune itself.
 
\begin{table*}
\caption{Physical and orbital properties of the planets of interest formed in scenario III. $a_{\text{i}}$ and $a_{\text{f}}$ are the initial and final semimajor axes in au, respectively, $M_{\text{i}}$ and $M_{\text{f}}$ the initial and final masses in $M_{\oplus}$, respectively, $T_{\text{GI}}$ is the timescale in Myr associated with the last giant impact produced by an embryo, and $W$ is the final percentage of water by mass after 200 Myr of evolution.}
\begin{center}
\begin{tabular}{|c|c|c|c|c|c|c|c|}
\hline
\hline
Simulation & $a_{\text{i}}$(au)  & $a_{\text{f}}$(au) & $M_{\text{i}}(M_{\oplus})$ & $M_{\text{f}}(M_{\oplus})$ & $T_{\text{GI}}$ (Myr) & $W(\%)$ \\
\hline
\hline
SIM 1  & 3 & 1.74 & 19.67 & 23.90 & 7.52 & 32 \\
\hline
SIM 2  & 3 & 1.47 & 19.67 & 25.10 & 4.63 & 26 \\
\hline
SIM 3  & 3 & 1.29 & 19.67 & 23.52 & 8x$\text{10}^{-4}$ & 35 \\
\hline
SIM 4  & 3 & 1.35 & 19.67 & 21.94 & 1.32 & 29 \\
\hline
SIM 5  & 3 & 1.35 & 19.67 & 31.74 & 151.12 & 32 \\
\hline
SIM 6  & 3 & 1.35 & 19.67 & 27.07 & 11.45 & 36 \\
\hline
SIM 7  & 3 & 1.35 & 19.67 & 23.77 & 2.44 & 33 \\
\hline
\end{tabular}
\end{center}
\label{tab:NepPlanets}
\end{table*}

{\bf General evolution of system:} In general terms, the overall progression of the system evolution can be described as follows. On the one hand, the embryos started very early to evolve through mutual collisions. Then, after $\sim$0.5 Myr, the number of embryos started to decrease dramatically, also as a result of collisions with the central star. At 10 Myr, $\sim$35\% of the embryos collided with the star and $\sim$30\% remained in the system by that time. At the end of the simulation, only $\sim$10\% of the embryos survived in the system. On the other hand, the planetesimals started their evolution by undergoing collisions with planetary embryos. However, Neptune's gravitational perturbations started to play an important role on the planetesimals after 1 Myr. At the end of the simulation, 50\% of the planetesimals were ejected from the system, while $\sim$40\% of them collided with the central star.

\begin{figure}[ht]
 \centering                    
 \includegraphics[angle=0, width= 0.45\textwidth]{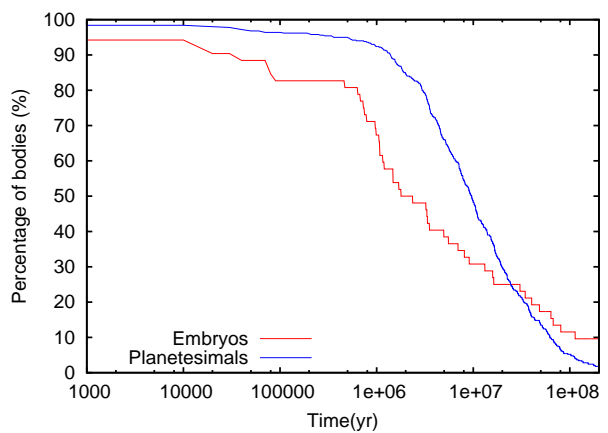}
 \caption{Percentage of planetary embryos (red curve) and planetesimals (blue curve) surviving in the system as a function of time for scenario III.}
\label{fig:NTN}     
\end{figure}

\begin{figure}[ht]         
 \centering         
 \includegraphics[angle=0, width= 0.45\textwidth]{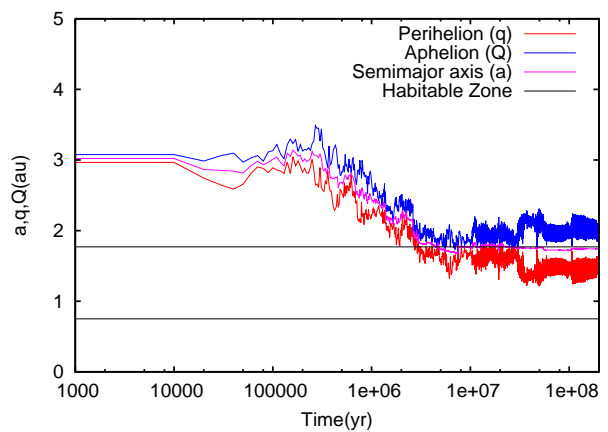} 
 \caption{Evolution in time of the semimajor axis and the perihelion and aphelion distances of the planet that survived in the HZ in SIM 4, corresponding to scenario III. The black lines indicate the limits of the optimistic HZ. A color version of this figure is available in the electronic version of the journal.}
\label{fig:OrbitNep}
\end{figure}

{\bf Orbital evolution:} When the gaseous phase ended, Neptune was located just beyond of the snow line at 3 au in all simulations. In particular, Figure~\ref{fig:OrbitNep} shows the evolution in time of the semimajor axis, and the perihelion and aphelion distances of Neptune in SIM 4, as representative of the whole group of simulations. In general terms, the inward migration of Neptune began at $\sim$0.1 Myr in all simulations, as a result
of interactions with the embryos located inside its initial orbit. Then, the planet reached the limits of the optimistic HZ between 4 Myr and 10 Myr. We remark that not all the orbits of the Neptunes are completely contained within the limits of the HZ. Figure~\ref{fig:OrbitNep}
shows that the aphelion of the orbit of Neptune resulting from SIM 4 is located outside the snow line during the whole evolution. 

{\bf Mass evolution:} The Neptunes began the simulations with a primordial mass of 19.7$M_{\oplus}$ and received a significant amount of impacts by planetary embryos and planetesimals over the 200 Myr. The maximum number of impacts that Neptune received by planetesimals and embryos in our simulations are 38 and 8, respectively. Most of the impactors are planetesimals. However, the truly significant contribution to Neptune's final mass comes
from the planetary embryos.

\begin{figure}[ht]            
  \centering
 \includegraphics[angle=0, width= 0.45\textwidth]{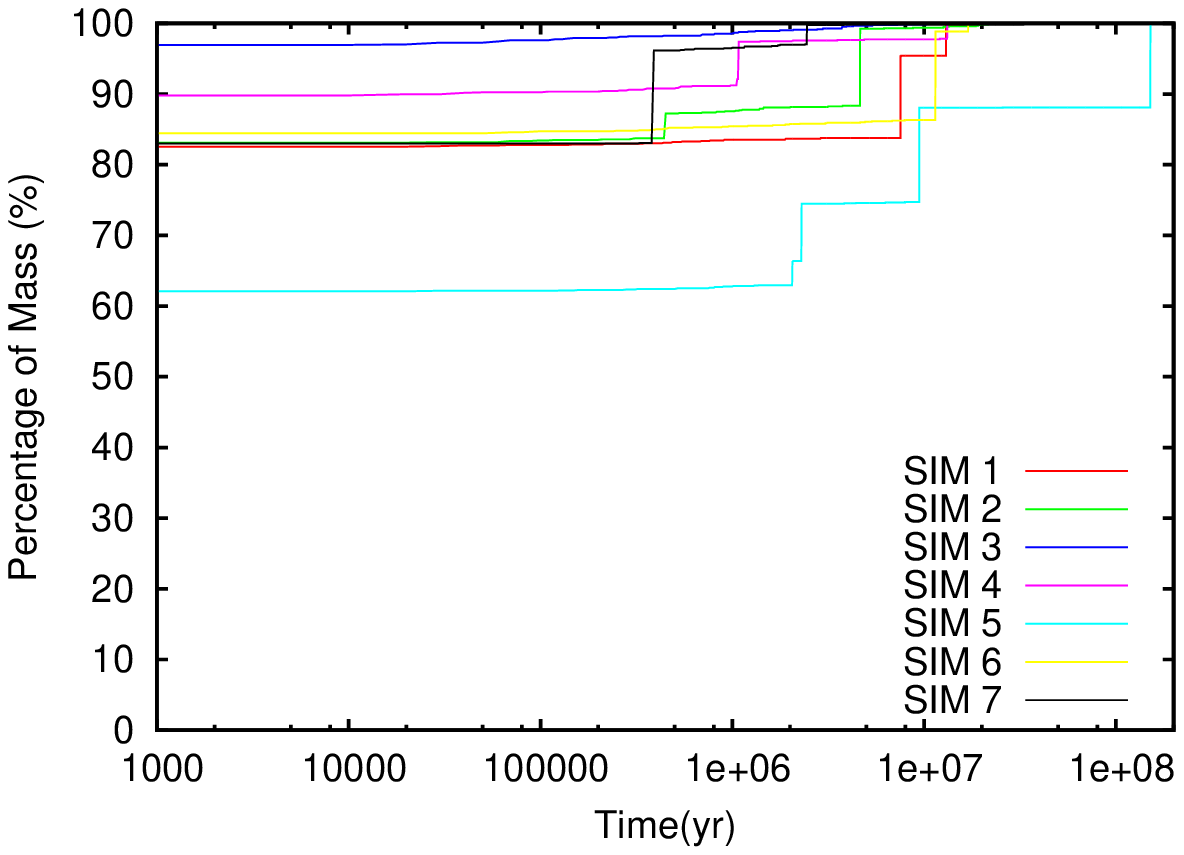}\\
 \includegraphics[angle=0, width= 0.45\textwidth]{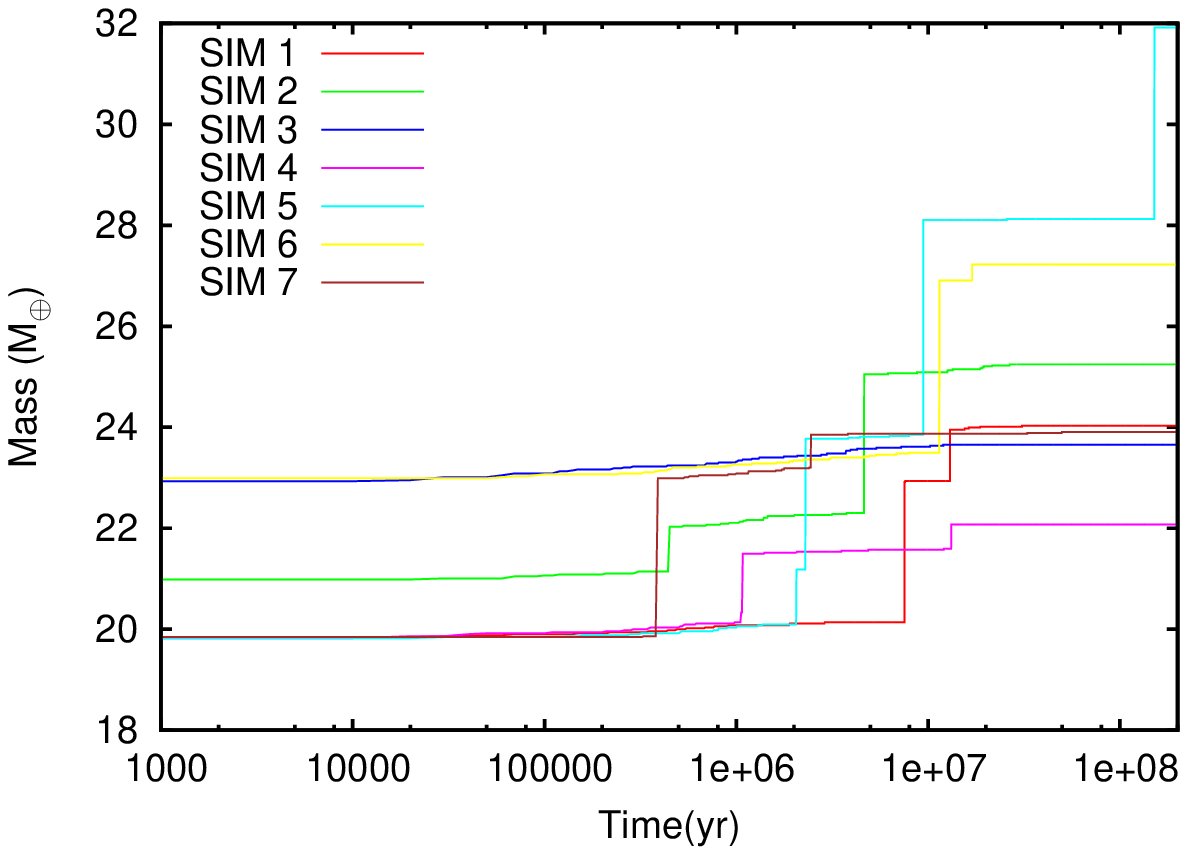}
 \caption{Percentage of mass (top) and total mass (bottom) as a function of time of the planets of interest in scenario III. A color version of this figure is available in the electronic version of the journal.}  
\label{fig:MassNep}
\end{figure}

Figure~\ref{fig:MassNep} shows the evolution in time of the mass (bottom panel) and percentage of mass (top panel) of the planets surviving in the HZ in this dynamical scenario. In general terms, collisional events occurred between 0.5 Myr and 11 Myr. However, since embryos and planetesimals were removed on slower timescales than in scenarios I and II, late impacts were also possible. In fact, the Neptunes resulting from SIMs 5 and 6 received their final impacts at $\sim$38 Myr and $\sim$151 Myr, respectively. Since the primordial mass of each Neptune is of $\sim$20$M_{\oplus}$, it is no surprise that at least 80\% of the planet's final mass is primordial in most simulations. From this point of view, collisions did not make a significant contribution to the final planetary mass. However, this is not the case for the Neptune that received late impacts in SIM 5. Figure~\ref{fig:MassNep} shows that the planet accreted up to 12 $M_{\oplus}$ in collisional events during the whole 200 Myr of evolution, meaning that collisions did make a significant contribution to the final planetary mass.

After 200 Myr of evolution, the Neptunes that survived in the HZ reached final masses between $\sim$22$M_{\oplus}$ and $\sim$32$M_{\oplus}$.

{\bf Mega-Earths or Neptunes?:} Our numerical simulations showed that the planet that survived within the limits of the HZ is Neptune. According to the semi-analytical model, this planet started its evolution during the post-gas phase with an initial mass of $\sim$20$M_{\oplus}$, of which 7.1$M_{\oplus}$ corresponds to the gaseous envelope. Our goal is not to discuss the possible astrobiological interest of such planets since some of their physical properties are not appropriate for the development and maintenance of life. However, it is probable that these objects lost a significant amount of their envelopes through collisions with planetary embryos during the dynamical evolution. Giant impacts on a planet cause global motion of the ground, creating a shock wave that can lead to a significant loss of the planetary atmosphere \citep{Schlichting2015}. If the atmospheric collisional remotion is relevant, these planets could be mega-Earths instead of Neptune-like planets. The mega-Earth concept was suggested from an observational point of view with the discovery of Kepler-10c, a planet with an upper limit on its mass of about 20$M_{\oplus}$ \citep{Fressin2011}. In order to answer this ambiguity, we need an $N$-body code that models the different collision regimes proposed by Leinhardt \& Stewart (2012) together with the considerations concerning the envelope removal proposed by \citet{Schlichting2015}. This task is not trivial and is beyond the scope of this present investigation. 

{\bf Water delivery}: Regardless of the true nature of the planets of interest in this scenario (mega-Earths or Neptunes), we are still interested in studying their final water content and how it was obtained during their evolution. These planets have a high primordial water content of ~6.5$M_{\oplus}$, which is half of their initial solid mass. Then, over the 200 Myr of evolution, they also accreted water in collisional events with water-rich bodies. As we discussed earlier, the maximum number of impacts with planetesimals is 38, and they were also hit by up to two water-rich embryos. These collisions made a significant contribution to the planets' final water content. In our simulations, the total mass of water accreted in such impacts reaches values up to $\sim$4$M_{\oplus}$. 

This dynamical scenario shows two possibilities for the planets that survived in the HZ, depending on the atmospheric evolution discussed earlier.

- If the planets that survived in the HZ suffered a significant atmospheric loss by giant impacts, they would end up being mega-Earths with up to 10$M_{\oplus}$ of water, with both primordial and collisional origins. However, these 10$M_{\oplus}$ of water must be interpreted as an upper limit, since atmospheric loss could also remove part of this water content. 

- If the envelopes were conserved, the planets that survived in the HZ would be giant Neptune-like planets with masses between $\sim$22$M_{\oplus}$ and $\sim$32$M_{\oplus}$, so we discard them as planets of interest. However, if the Neptunes were able to form regular satellites during their evolution, the 4$M_{\oplus}$ of collisional water could have been available for accretion by those satellites. In this case, the formation of habitable moons could be possible. This process is not the topic of our investigation, but further studies could very well determine the astrobiological interest of this scenario.

\subsection{Scenario IV: Super-Earth I}

In the present scenario, we study the formation of planets in the HZ and the water delivery process in systems whose most massive planet is a super-Earth of $\sim$5$M_{\oplus}$. This planet is located around the snow line at the end of the gaseous phase and acts as the main perturber of the system. Because of the stochastic nature of the accretion process and the high CPU time required to carry out the numerical simulations, we performed six $N$-body simulations of planetary accretion. Of these six numerical simulations, five formed one planet in the HZ after 200 Myr of evolution. It is worth noting that the accretion seed of four of the five planets that survived in the HZ is the 5$M_{\oplus}$ super-Earth formed at 3 au at the end of the gaseous phase. Table~{\ref{tab:SEIPlanets}} summarizes the main physical properties and orbital parameters of the planets that survived in the HZ.

\begin{table*}
\caption{Physical and orbital properties of the planets of interest formed in scenario IV. $a_{\text{i}}$ and $a_{\text{f}}$ are the initial and final semimajor axes in au, respectively, $M_{\text{i}}$ and $M_{\text{f}}$ the initial and final masses in $M_{\oplus}$, respectively, $T_{\text{GI}}$ is the timescale in Myr associated with the last giant impact produced by an embryo, and $W$ is the final percentage of water by mass after 200 Myr of evolution.}
\begin{center}
\begin{tabular}{|c|c|c|c|c|c|c|c|}
\hline
\hline
Simulation & $a_{\text{i}}$(au)  & $a_{\text{f}}$(au) & $M_{\text{i}}(M_{\oplus})$ & $M_{\text{f}}(M_{\oplus})$ & $T_{\text{GI}}$ (Myr) & $W(\%)$ \\
\hline
\hline
SIM 1  & 3 & 1.30 & 4.94 & 8.29 & 5.53 & 40 \\
\hline
SIM 2  & 3.48 & 1.24 & 1.43 & 7.65 & 40.22 & 27 \\
\hline
SIM 3  & 3 & 1.66 & 4.94 & 9.14 & 43.62 & 35 \\
\hline
SIM 4  & 3 & 1.54 & 4.94 & 12.00 & 183 & 32 \\
\hline
SIM 5  & 3 & 1.68 & 4.94 & 8.50 & 41.27 & 40 \\
\hline
\end{tabular}
\end{center}
\label{tab:SEIPlanets}
\end{table*}

\begin{figure*}[ht]
 \centering
 \includegraphics[angle=0, width= 0.45\textwidth]{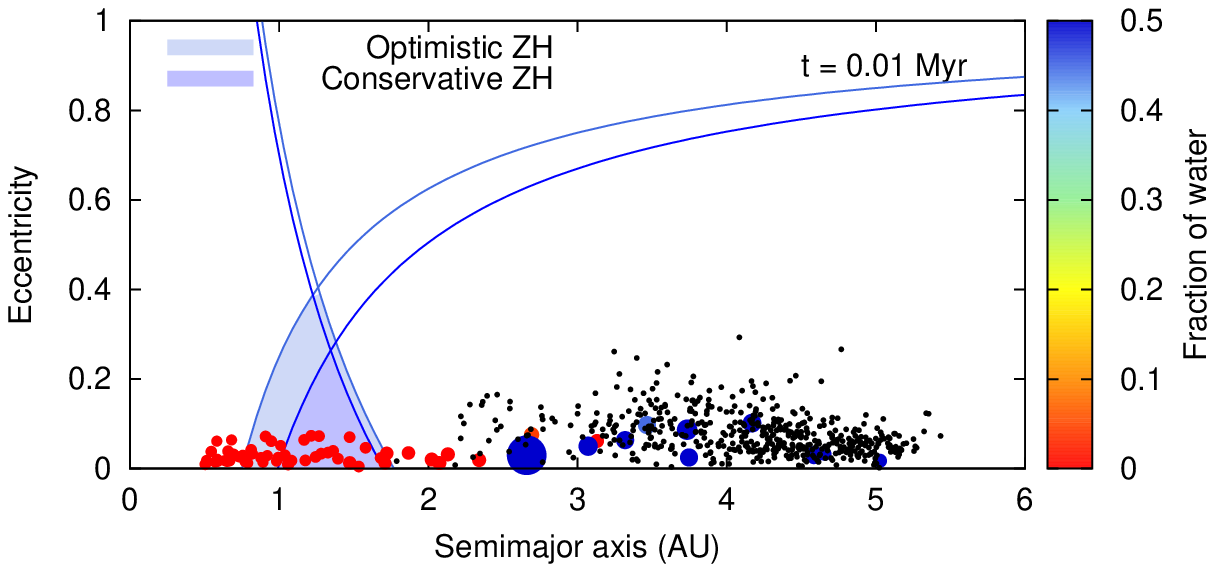}
 \includegraphics[angle=0, width= 0.45\textwidth]{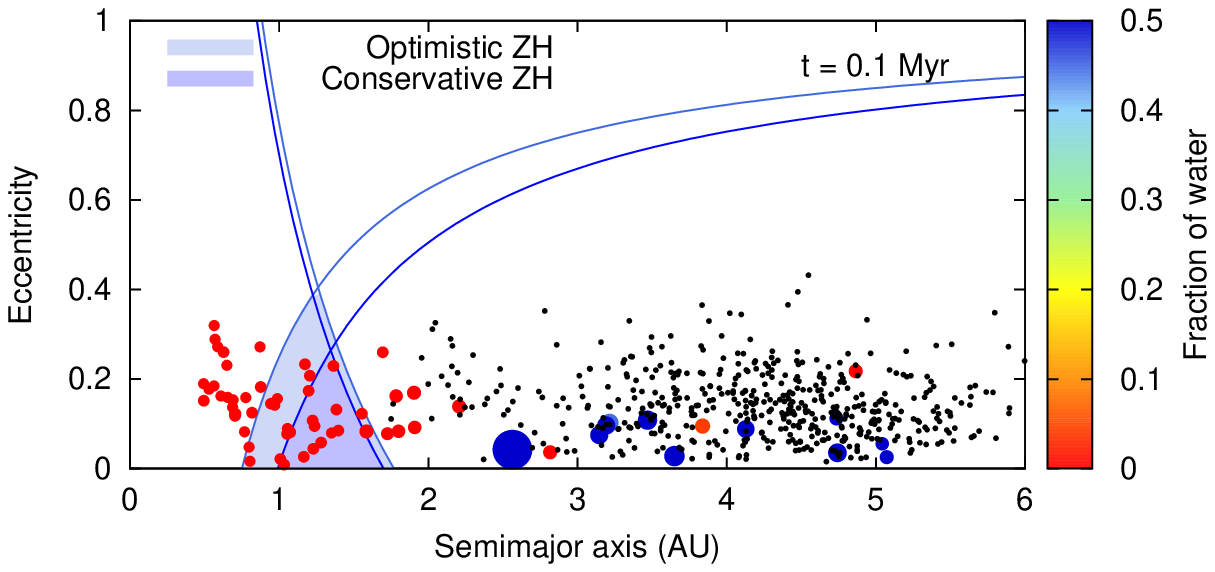}\\
 \includegraphics[angle=0, width= 0.45\textwidth]{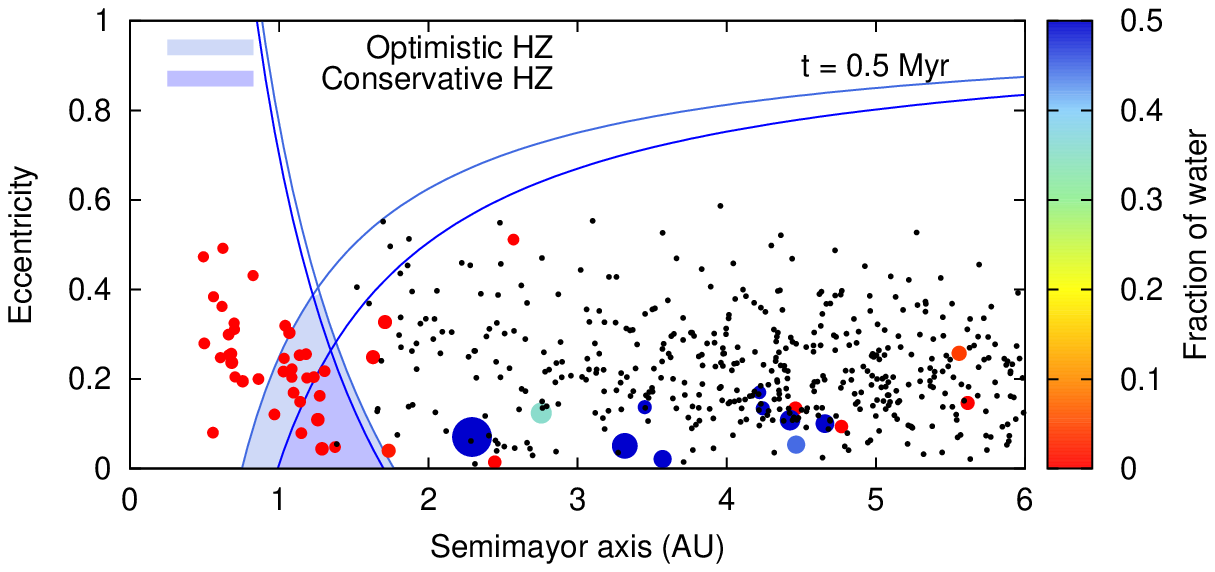}
 \includegraphics[angle=0, width= 0.45\textwidth]{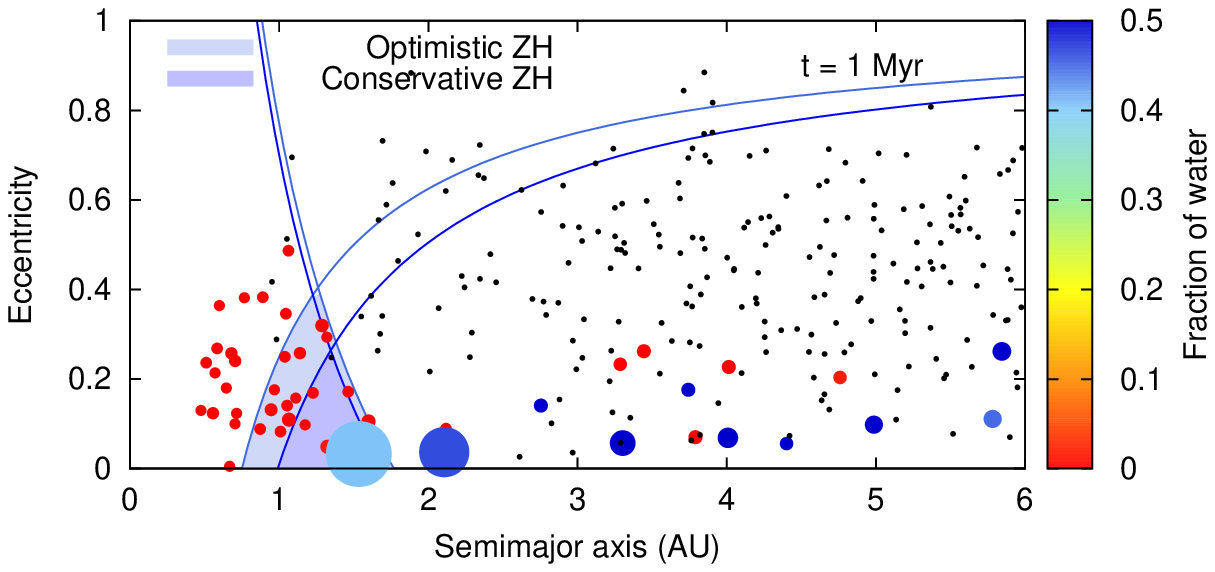}\\
 \includegraphics[angle=0, width= 0.45\textwidth]{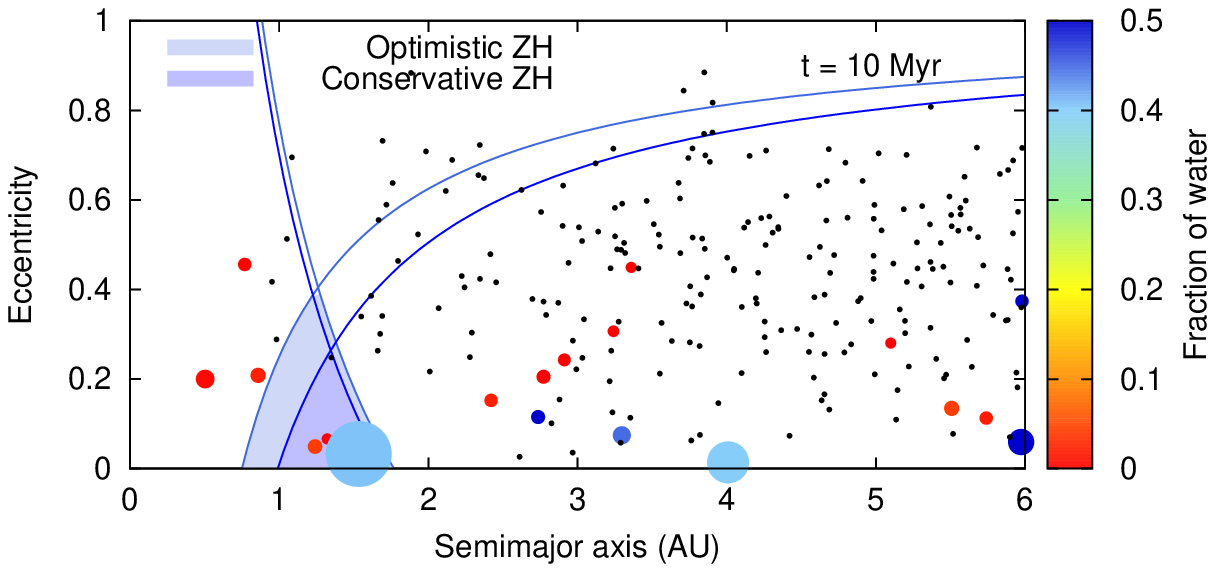}
 \includegraphics[angle=0, width= 0.45\textwidth]{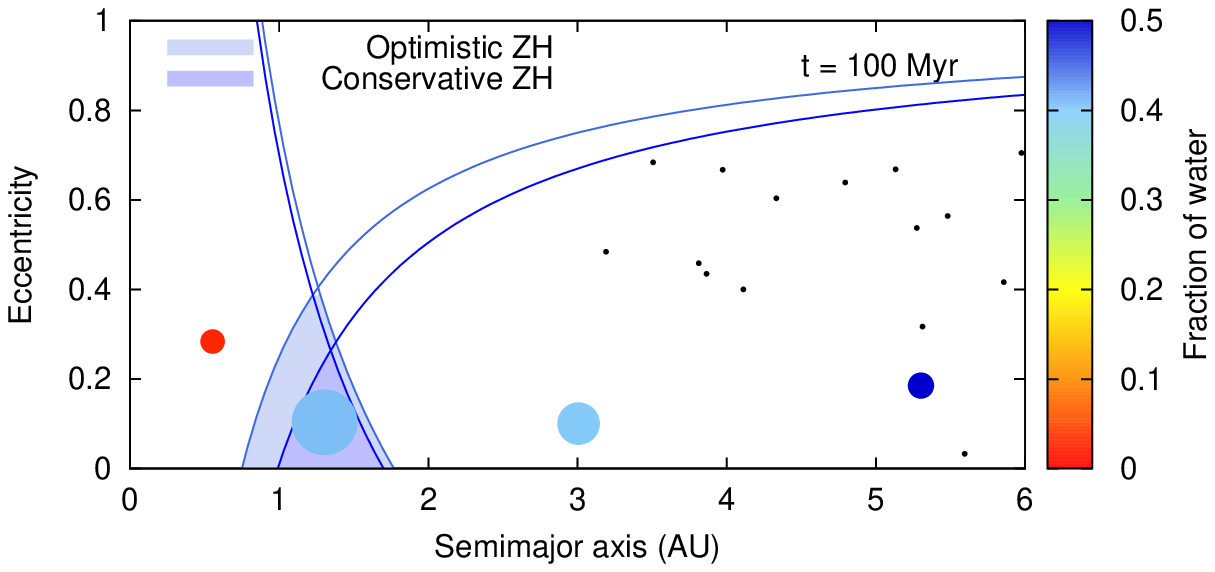}
 \caption{Evolution in time of SIM 1 corresponding to scenario IV. Planetary embryos are plotted as colored circles following the color scale that represents the fraction of water of the embryos with respect to their total masses. The blue and light
blue shaded areas represent the conservative and the optimistic HZ, respectively. Moreover, the blue and light blue curves represent curves of constant perihelion and aphelion, both for the conservative and the optimistic HZ. At the end of the simulation, a super-Earth of 8.29$M_{\oplus}$ and 40\% water content survives within the limits of the HZ. It is worth noting that the accretion seed of this planet is the 5$M_{\oplus}$ super-Earth formed at 3 au at the end of the gaseous phase. A color version of this figure is only available in the electronic version.}
\label{fig:SEISnapshot}
\end{figure*}

\begin{figure}[ht]
 \centering                    
 \includegraphics[angle=0, width= 0.45\textwidth]{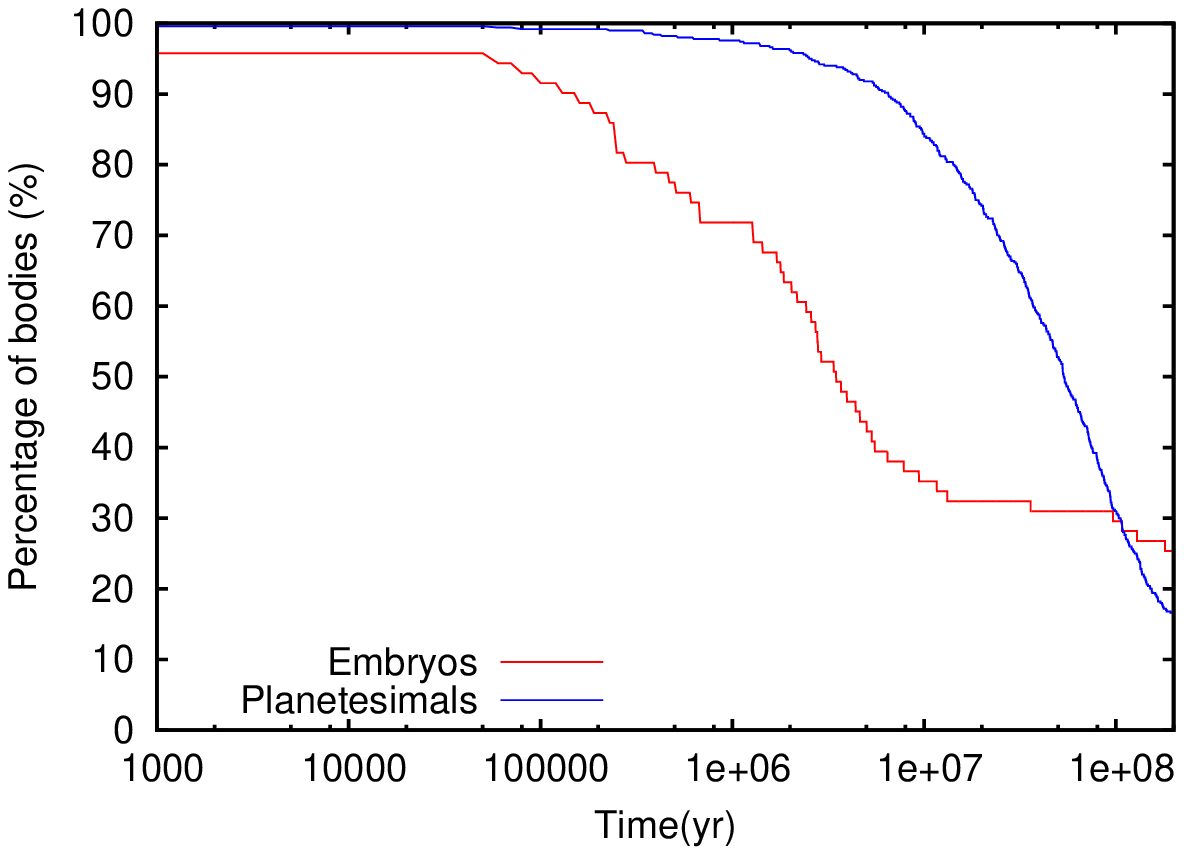} 
 \caption{Percentage of planetary embryos (red curve) and planetesimals (blue curve) remaining in the system as a function of time for scenario IV. A color version of this figure is only available in the electronic version.}                                                                               
\label{fig:NTst}
\end{figure}

{\bf General evolution of system:} The dynamical evolution of the simulations that produce planets in the HZ is very similar. We therefore analyze the results of SIM 1 as representative of the whole group. Figure~\ref{fig:SEISnapshot} shows six snapshots of the evolution in time on the semimajor axis-eccentricity plane of SIM 1. The 5$M_{\oplus}$ super-Earth located at 3 au, which initially represents the main perturber of the system, is not massive enough to efficiently excite the distributions of embryos and planetesimals as Jupiter and Saturn did. Figure~\ref{fig:NTst} shows the fraction of embryos and planetesimals remaining in the system as a function of time. The system evolved on slower timescales than in scenarios I and II. On the one hand, planetary embryos only evolved through collisions with planetesimals and other embryos. The collisional events between embryos occurred mostly between $\sim$0.1 Myr and $\sim$10 Myr, after which fewer than 30\% of embryos survived in the system. On the other hand, the collisions between planetesimals and embryos started at $\sim$1 Myr. Then, after $\sim$10 Myr, planetesimals began to be ejected from the system and collided with the central star. At the end of the simulation, fewer than 20\% of planetesimals survived in the system. After 200 Myr of evolution, the planetary system resulting from SIM 1 is composed of four planets with semimajor axes smaller than 5 au, one of which survives in the HZ: a super-Earth with a mass of 8.29$M_{\oplus}$ and 40\% of water by mass.


\begin{figure}[ht]         
 \centering                    
 \includegraphics[angle=0, width= 0.45\textwidth]{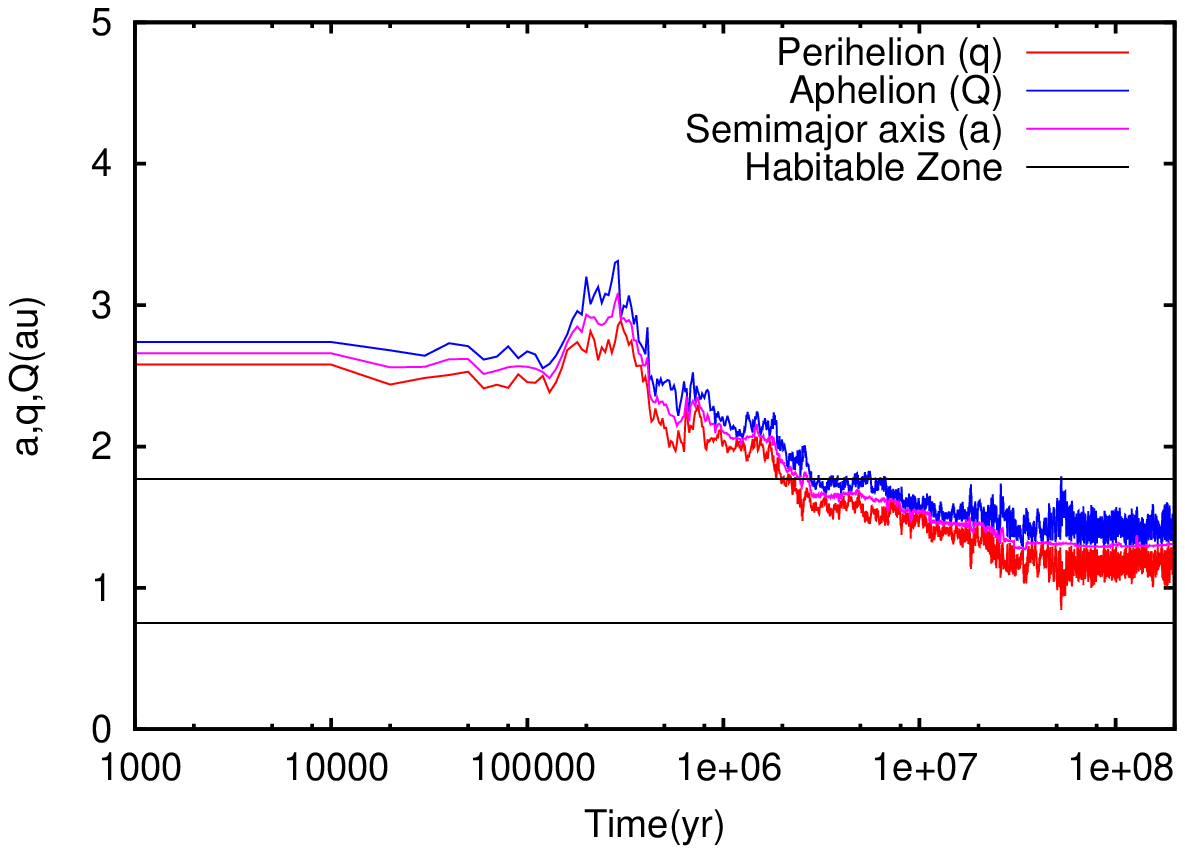}\\
 \caption{Evolution in time of the semimajor axis $a$, and the perihelion $q$ and aphelion $Q$ distances of the super-Earth that survived in the HZ in SIM 1 corresponding to scenario IV. The black lines indicate the limits of the optimistic HZ. A color version of this figure is only available in the electronic version.}
\label{fig:OrbitSEI}
\end{figure}

{\bf Orbital evolution:} Table~{\ref{tab:SEIPlanets}} shows that in four of five numerical simulations, the planet that survived in the HZ is the 5$M_{\oplus}$ super-Earth that was formed at 3 au at the end of the gaseous phase, while the planet that survived in SIM 2 evolved from an accretion seed initially located at 3.48 au. In particular, Figure~\ref{fig:OrbitSEI} shows the evolution in time of the semimajor axis and the perihelion and aphelion distances of the planet of interest resulting from SIM 1 as representative of the whole group. The migration of the super-Earth, due to close encounters with the inner embryos, began at $\sim$0.1 Myr and it reached the limits of the HZ in $\sim$2 Myr. Then, the planet remained in that region until the end of the 200 Myr evolution. We find that in general terms, the other planets of interest formed in this scenario evolved in a similar way. However, the planets formed in SIMs 3-5 reached aphelion distances larger than the outer limit of the optimistic HZ. We remark that these planets in SIMs 3, 4, and 5 reached semimajor axes and maximum eccentricities of 1.66, 0.25, and 1.54 au and 0.43, 1.68, and 0.33, respectively. We consider them as planets of interest following the criterion given by \citet{Williams2002}.

\begin{figure}[ht]            
 \centering
 \includegraphics[angle=0, width= 0.45\textwidth]{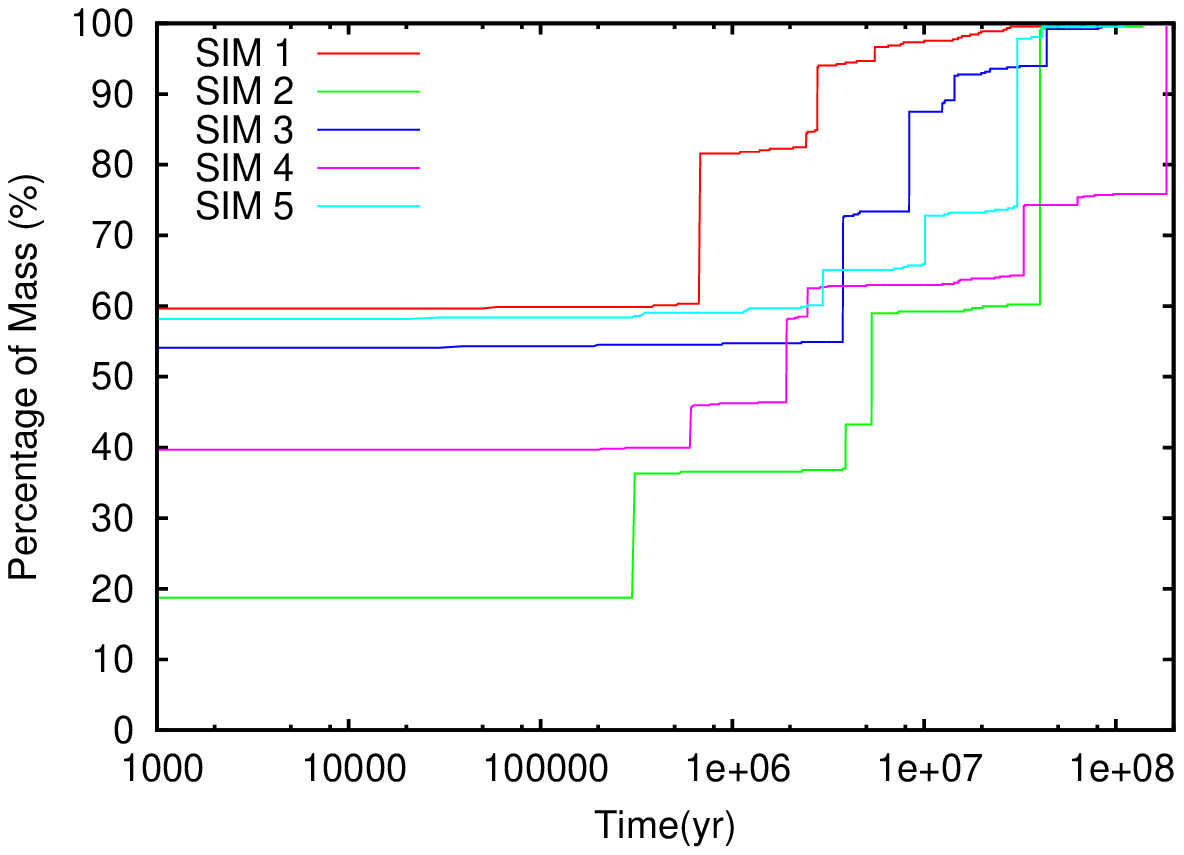}\\
 \includegraphics[angle=0, width= 0.45\textwidth]{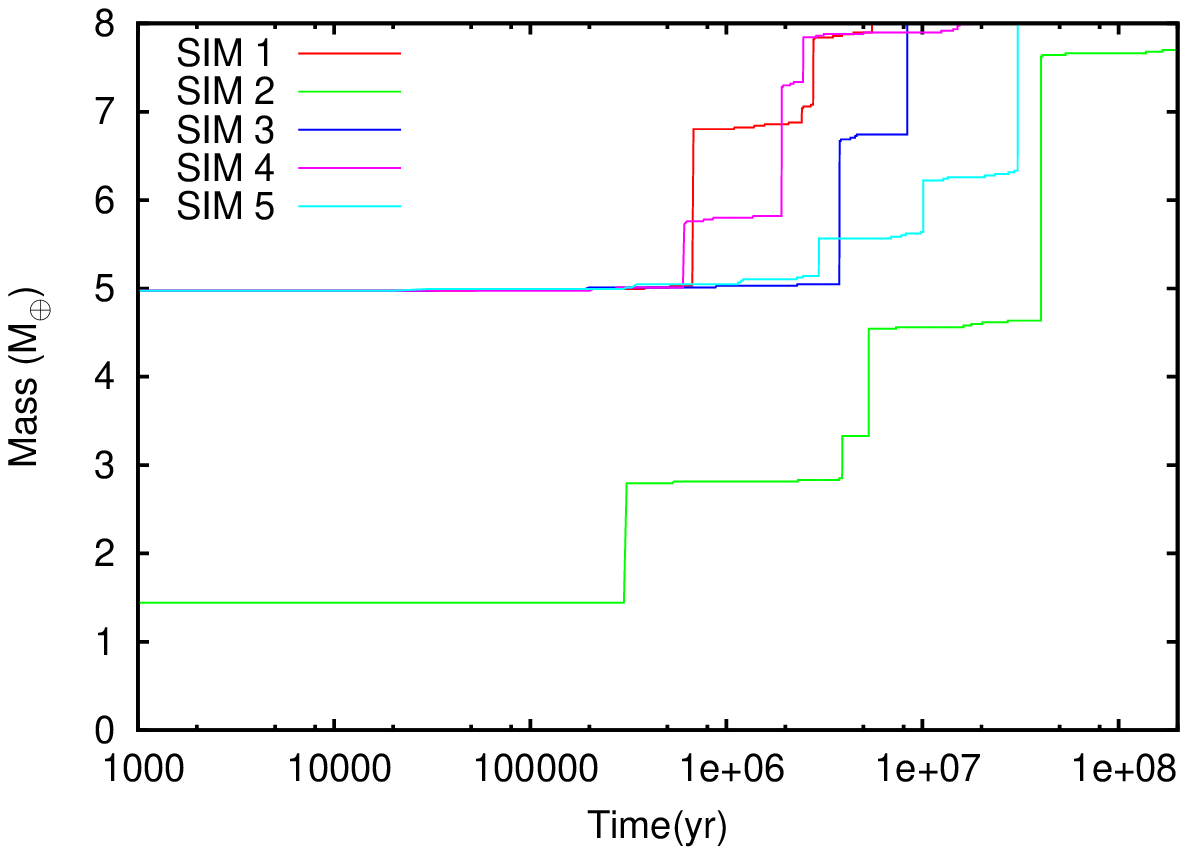}
 \caption{Percentage of mass (top) and total mass (bottom) as a function of time of the planets of interest in scenario IV. A color version of this figure is available in the electronic version of the journal.}  
\label{fig:MassSEI}
\end{figure}

{\bf Mass evolution:} Figure~\ref{fig:MassSEI} shows the evolution in time of the masses (bottom panel) and percentages of mass (top panel) of the planets surviving in the HZ in this scenario. These planets have primordial masses of $\sim$5$M_{\oplus}$ (SIMs 1, 3, 4, and 5) and 1.43$M_{\oplus}$ (SIM 2). Then, during the 200 Myr of evolution, they were hit by many embryos and planetesimals. The maximum number of impacts that the planets had with planetesimals and planetary embryos were 28 and 6, respectively. However, we remark that the impactors that significantly increased the mass of each of those planets of interest were the planetary embryos. In this scenario, collisions made a very important contribution to the mass of the planets. Figure~\ref{fig:MassSEI} shows that between 40\% and 80\% of their final mass was accreted by collisions. 

In general terms, the planets that survived in the HZ required formation timescales shorter than $\sim$44 Myr, since the giant collisional events occurred mostly between $\sim$0.5 Myr and $\sim$44 Myr. These timescales are shorter than the timescale associated with Earth \citep{Jacobson2014}. However, since the remotion of bodies in this scenario was a slow process, as Figure~\ref{fig:NTst} shows, late giant impacts were indeed possible. The planet that survived in the HZ in SIM 4 received a giant impact at 180 Myr, which led to a formation timescale much longer than the timescale associated with Earth.

After 200 Myr of evolution, the planets that survived in the HZ in the present scenario are very massive super-Earths with masses between 7.65$M_{\oplus}$ and 12$M_{\oplus}$.

\begin{figure*}[ht]
 \centering
 \includegraphics[angle=0, width= 0.45\textwidth]{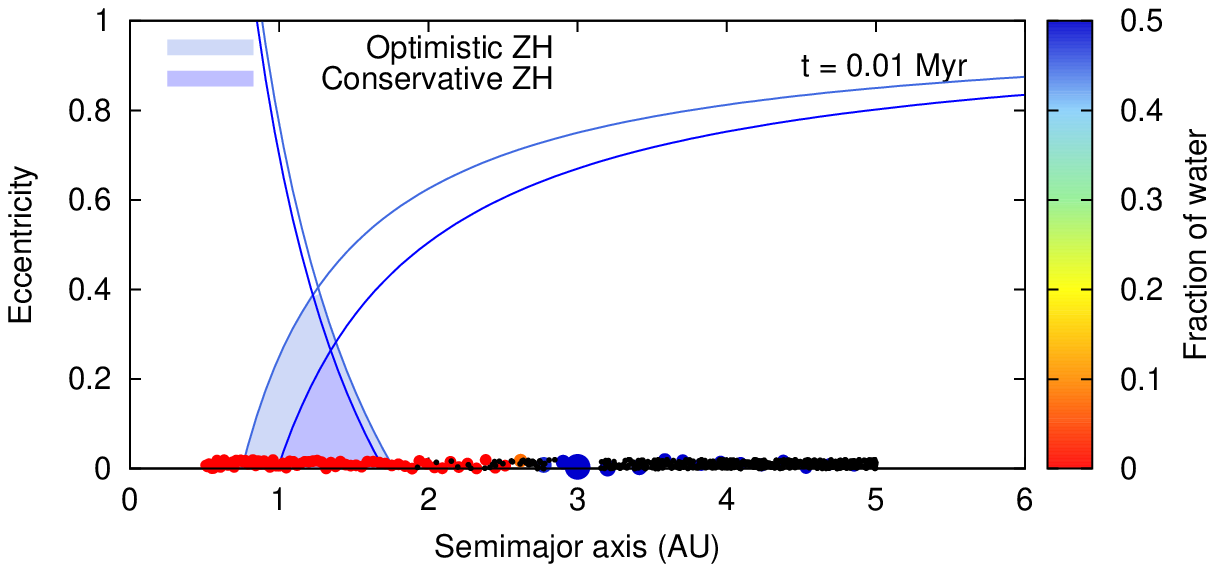}
 \includegraphics[angle=0, width= 0.45\textwidth]{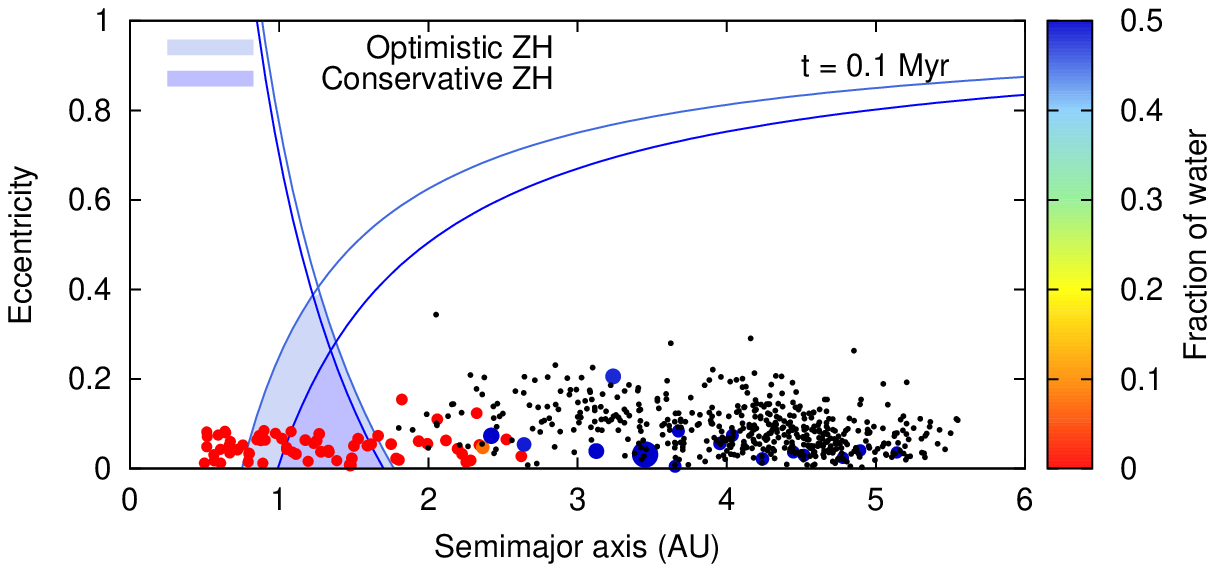}\\
 \includegraphics[angle=0, width= 0.45\textwidth]{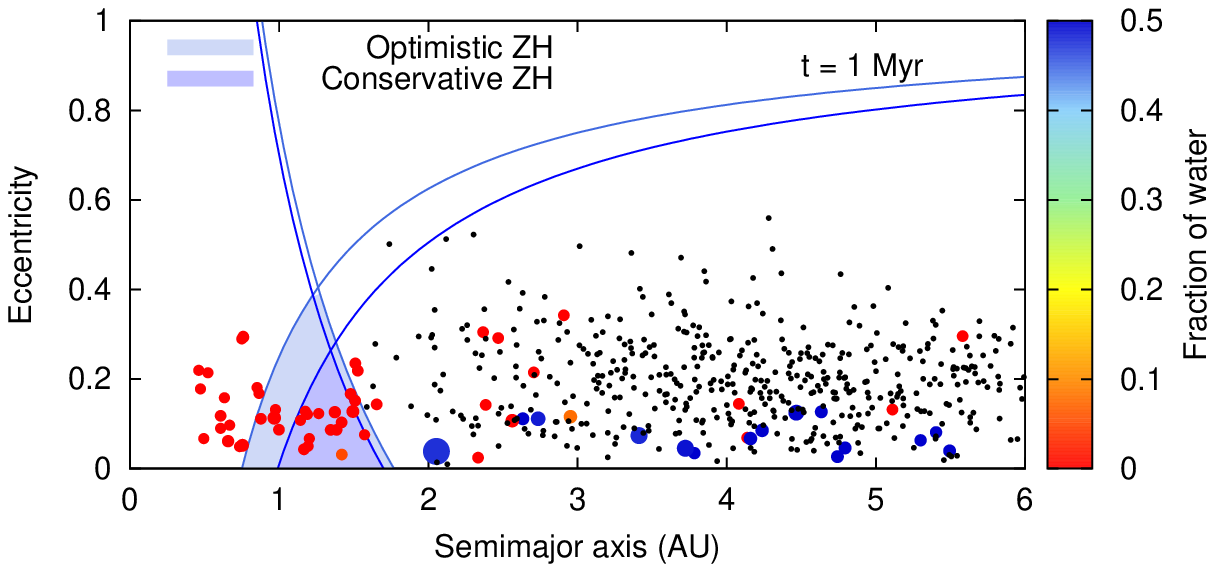}
 \includegraphics[angle=0, width= 0.45\textwidth]{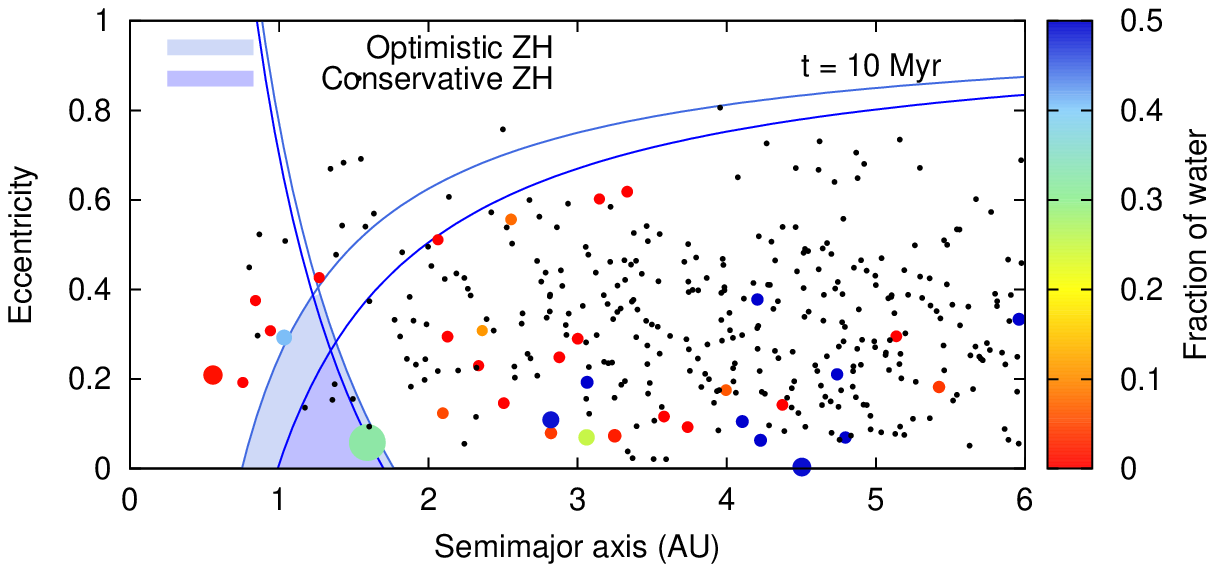}\\
 \includegraphics[angle=0, width= 0.45\textwidth]{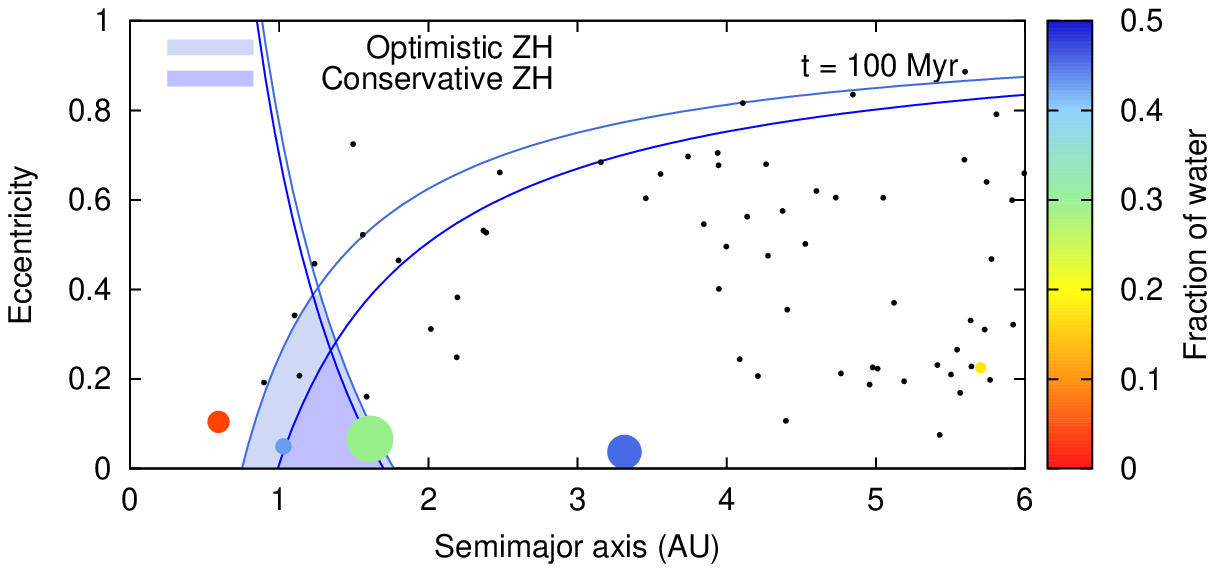}
 \includegraphics[angle=0, width= 0.45\textwidth]{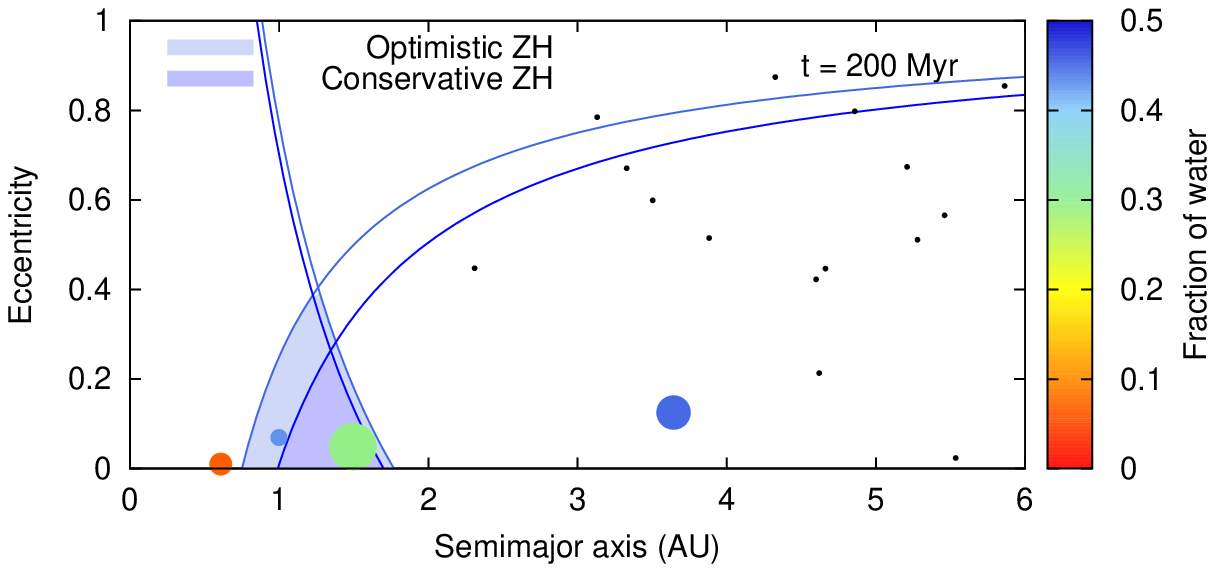}
 \caption{Evolution in time of SIM 5 corresponding to scenario V. Planetary embryos are plotted as colored circles following the color scale that represents the fraction of water respect to their total masses. The blue and light blue shaded areas represent the conservative and the optimistic HZ, respectively. Moreover, the blue and light blue curves represent curves of constant perihelion and aphelion for the conservative and optimistic HZ. At the end of the simulation, two planets of 1.05$M_{\oplus}$ and 6.21$M_{\oplus}$ survive within the limits of the HZ with 42\% and 29\% of water by mass, respectively. A color version of this figure is only available in the electronic version.}
\label{fig:SEIISnapshot}
\end{figure*}

{\bf Water delivery:} We proceed to analyze the water accretion of the planets that survived in the HZ. The five planets of interest evolved from accretion seeds that formed outside the snow line during the gas phase, which means that half of their primordial masses are composed of water. Unlike all water worlds formed in scenario I and most of the worlds produced in scenario II, the planets that survived in the HZ in this scenario continued to accrete water through impacts with planetary embryos and planetesimals during the post-gas phase. In general terms, the process of water accretion occurred on timescales of a few tenths of Myr.

The maximum amount of water accreted by collisions with planetary embryos and planetesimals was 1.33$M_{\oplus}$ and 0.25$M_{\oplus}$, respectively. Clearly, most of the collisional water was provided by embryos. The total water mass of the planets that survived in the HZ is between 2.12$M_{\oplus}$ and 4.06$M_{\oplus}$. This
means that impacts with water-rich bodies made an important contribution to the final water contents of the planets. Between 22\%-63\% of the total water content was accreted by collisions, while the rest is primordial. We find that the planets that survived in the HZ in this scenario are truly water worlds since they evolved from accretion seeds initially located beyond the snow line and ended with 27\%-40\% of water by mass. \\

The results described in the present section suggest that the formation of planets in the HZ is a very efficient process in this dynamical scenario. Five of six $N$-body simulations produced planets in the HZ, each of which is a water world. Like in the previous scenarios, the systems harboring a 5$M_{\oplus}$ as main perturber were not able to form planets with physical and dynamical properties similar to the Earth.

\subsection{Scenario V: Super-Earth II}

Here, we present the results of our last dynamical scenario, which is defined by a planetary system without gas giants and a super-Earth of $\sim$2.5$M_{\oplus}$ at 3 au acting as main perturber. Just as in scenario IV, we performed only six $N$-body simulations because of the high CPU time required. Table~{\ref{tab:SEIIPlanets}} summarizes the main physical properties and orbital parameters of all planets formed in the HZ in the present scenario. We find that four simulations formed one planet in the HZ, while two of them produced two planets in such a region. We remark that this planetary configuration with two planets coexisting in the HZ was only achieved in this dynamical scenario.

\begin{table*}
\caption{Physical and orbital properties of the planets of interest formed in Scenario V. $a_{\text{i}}$ and $a_{\text{f}}$ are the initial and final semimajor axes in au, respectively, $M_{\text{i}}$ and $M_{\text{f}}$ the initial and final masses in $M_{\oplus}$, respectively, $T_{\text{GI}}$ is the timescale in Myr associated with the last giant impact produced by an embryo, and $W$ is the final percentage of water by mass after 200 Myr of evolution.}
\begin{center}
\begin{tabular}{|c|c|c|c|c|c|c|}
\hline
\hline
Simulation & $a_{\text{i}}$(au) & $a_{\text{f}}$(au) & $M_{\text{i}}(M_{\oplus})$ & $M_{\text{f}}(M_{\oplus})$ & $T_{\text{GI}}$ (Myr) & $W(\%)$ \\
\hline
\hline
SIM 1  & 1.67 & 0.92 & 0.14 & 3.40 & 34.09 & 15 \\
\hline
SIM 2  & 3 & 1.54 & 2.56 & 5.56 & 31 & 34 \\
\hline
SIM 3  & 3 & 1.54 & 2.56 & 5.43 & 28.09 & 42 \\
\hline
SIM 4  & 3 & 1.59 & 2.56 & 6.60 & 18.88 & 35 \\
\hline
SIM 5  & 2.9 & 0.99 & 0.64 & 1.05 & 7.23 & 42 \\
   & 3   & 1.49 & 2.56 & 6.21 & 43.07 & 29 \\
\hline
SIM 6  & 1.89 & 1.12 & 0.17 & 1.25 & 9.32 & 6 \\
   & 3    & 1.73 & 2.56 & 5.89 & 110.98 &  42 \\
\hline\end{tabular}
\end{center}
\label{tab:SEIIPlanets}
\end{table*}

\begin{figure}[ht]
 \centering                    
 \includegraphics[angle=0, width= 0.45\textwidth]{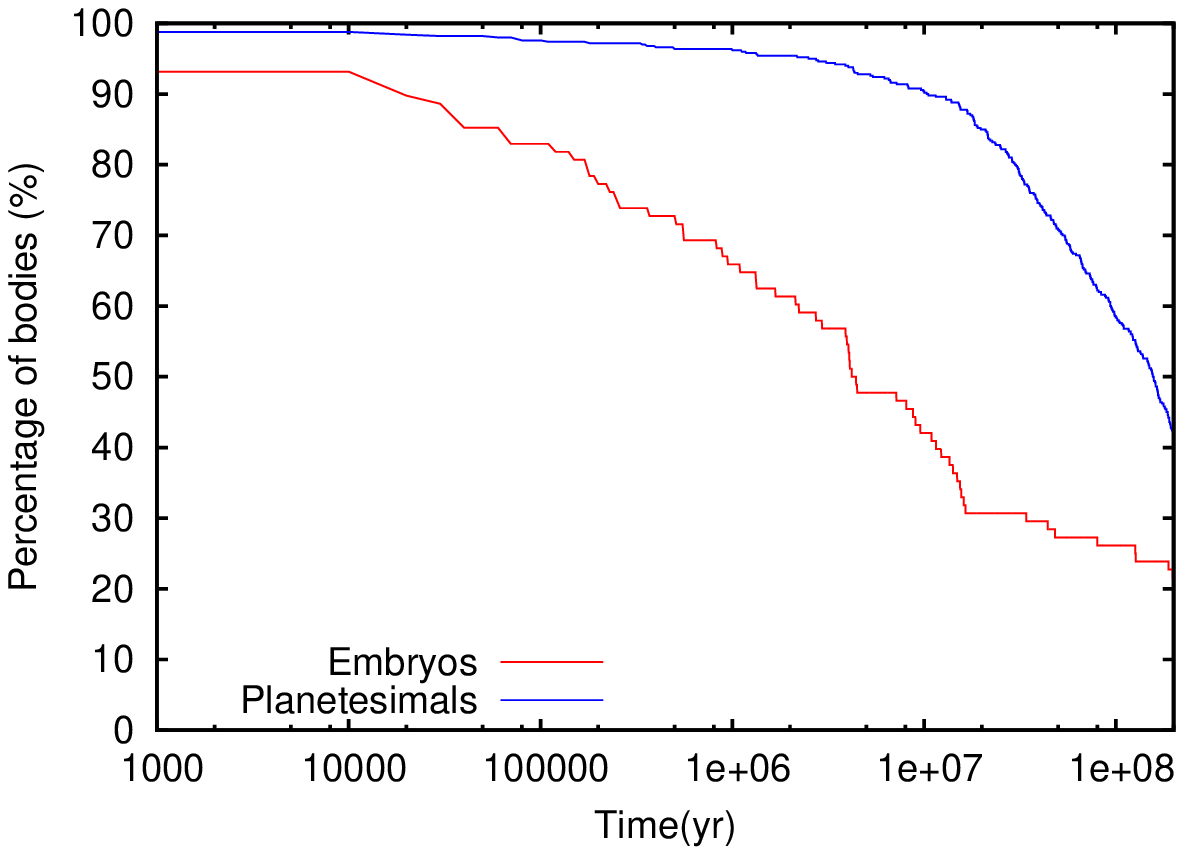} 
 \caption{Percentage of planetary embryos (red curve) and planetesimals (blue curve) remaining in the the system as a function of time for scenario V. A color version of this figure is available in the electronic version of the journal.}
\label{fig:NTSEII}
\end{figure}

{\bf General evolution of system:} Since the general evolution of all systems is very similar, we describe the evolution of SIM 5 as representative of the whole group of simulations. Figure~\ref{fig:SEIISnapshot} shows six snapshots in time of the system on the semimajor axis-eccentricity plane. A super-Earth of $\sim$2.5$M_{\oplus}$ is the main perturber of the system, but this is not massive enough to gravitationally excite the embryos and planetesimals in a significant way. On the one hand, the embryo population only evolved through mutual collisions with other embryos. After 200 Myr, there were no ejections from the system nor collisions with the central star by planetary embryos. On the other hand, we find that planetesimals mostly impacted with embryos, but $\sim$40\% were ejected or collided with the star.  

Figure~\ref{fig:NTSEII} shows the fraction of embryos and planetesimals surviving in the system as a function of time. We find that the general evolution of the systems in the present dynamical scenario is even slower than in scenario IV. Fewer than half of the planetary embryos remained in the system at 10 Myr and $\sim$20\% survived at the end of the simulation. The dynamical evolution is even slower for the planetesimals. $\text{About }$90\% of planetesimals remained in the system at 10 Myr, while $\sim$40\% of them survived in the system at 200 Myr.

Figure~\ref{fig:SEIISnapshot} shows that the planetary system acquired its final configuration between 100 Myr and 200 Myr. The final planetary system is composed of four planets with semimajor axes smaller than 5 au, of which two survived in the HZ: a super-Earth with a final mass of 6.21$M_{\oplus}$ and 29\% of water by mass, and an Earth-mass planet with a final mass of 1.05$M_{\oplus}$ and 42\% of water by mass.

\begin{figure}[ht]         
 \centering                    
 \includegraphics[angle=0, width= 0.45\textwidth]{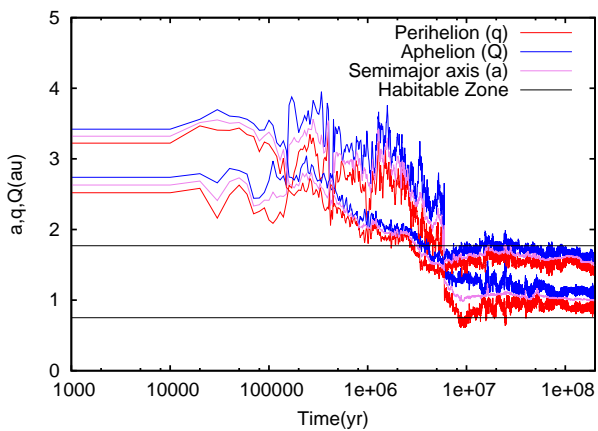}
 \caption{Evolution in time of the semimajor axis $a$, and the perihelion $q$ and aphelion $Q$ distances for the two planets of interest in SIM 5, corresponding to scenario V. The black lines indicate the limits of the optimistic HZ. A color version of this figure is available in the electronic version of the journal.}
\label{fig:OrbitSEII}
\end{figure}

{\bf Orbital evolution}: Table~{\ref{tab:SEIIPlanets}} lists the initial semimajor axis of the planets of interest formed in this scenario. The planets that survived in the HZ in this dynamical scenario evolved from different types of accretion seeds, depending on their starting position at the end of the gas phase.
\begin{itemize}
\item {\it Inside the snow line:} The planet of interest resulting from SIM 1 and the inner planet that survived in the HZ in SIM 6 evolved from accretion seeds that were located inside the snow line at the end of the gas phase. We find that the planet in SIM 1 remained in the HZ during the whole simulation, while the inner planet in SIM 6 was located slightly outside the HZ and entered that region in $\sim$0.2 Myr.
\item {\it Outside the snow line:} The planets that survived in the HZ in the rest of the simulations evolved from accretion seeds that were located beyond the snow line at the end of the gas phase. Then, they migrated inward into the HZ through interactions with the inner embryos. In particular, Figure~\ref{fig:OrbitSEII} shows the evolution in time of the semimajor axis, and the perihelion and aphelion distances of the two planets that survived in the HZ in SIM 5. The inward migration of the 6.21$M_{\oplus}$ super-Earth and the 1.05$M_{\oplus}$ inner planet began at $\sim$0.1 Myr and $\sim$1 Myr, respectively, and both reached the limits of the HZ in less than 10 Myr. Afterward, the orbits of both planets were fully contained inside the HZ until the end of the simulation. The orbits of the rest of the planets of interest formed in this scenario evolved in a similar way as the super-Earth that survived in the HZ in SIM 5. However, the planet formed in SIM 4 and the super-Earth in SIM 6 have aphelion distances larger than the outer limit of the optimistic HZ. These planets have maximum eccentricities of 0.25 and 0.17, respectively. Following the criterion given by \citet{Williams2002}, we find that these two planets are planets of interest. 
\end{itemize}

\begin{figure}[ht]            
 \centering
  \includegraphics[angle=0, width= 0.45\textwidth]{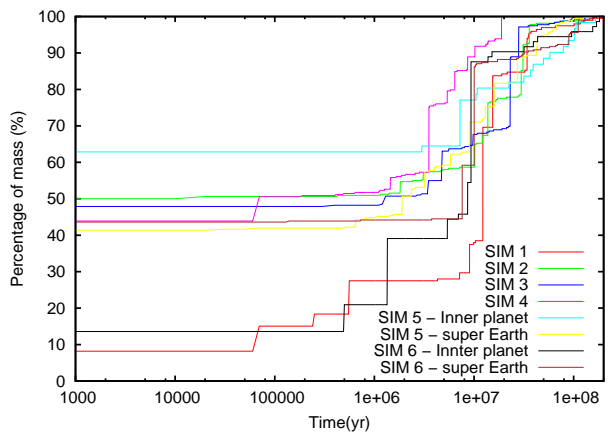}\\
  \includegraphics[angle=0, width= 0.45\textwidth]{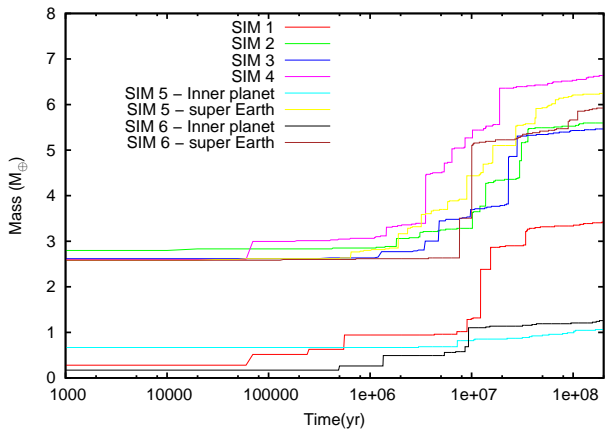}
 \caption{Percentage of mass (top) and total mass (bottom) as a function of time of the planets of interest in scenario V. A color version of this figure is available in the electronic version of the journal.}  
\label{fig:MassSEII}
\end{figure}

{\bf Mass evolution:} Figure~\ref{fig:MassSEII} shows the evolution in time of the planetary mass (bottom panel) and percentage of mass (top panel) of all the planets of interest that formed in this scenario. These planets have primordial masses ranging between 0.14$M_{\oplus}$ and 2.56$M_{\oplus}$. Then, they received many impacts from both planetary embryos and planetesimals during the 200 Myr of evolution. The maximum number of impacts with embryos and planetesimals was 13 and 44, respectively. In general terms, we find that the mass accreted through impacts was provided by both embryos and planetesimals in a similar way. In general terms, Figure~\ref{fig:MassSEII} shows that $\sim$35\%-60\% of the planetary masses was accreted by collisions. However, the super-Earth formed in SIM 1 and the inner planet produced in SIM 6 received $\sim$90\% of their final planetary masses by impacts. The collisional contribution to the planetary masses is therefore very important in this dynamical scenario.  

In general terms, the planets of interest required formation timescales of up to $\sim$40 Myr, which were determined by the time of the last giant impact with an embryo. These timescales are shorter than the timescale associated with the formation of Earth, as was discussed in previous sections. However, as in scenarios III and IV, late collisions were also possible because
of the slow body-removal timescales. The 5.89$M_{\oplus}$ super-Earth that survived in the HZ in SIM 6 received its last giant impact at 111 Myr, which represents a formation timescale that is similar to the formation timescale associated with Earth.

After 200 Myr of evolution, the six systems under study formed two Earth-mass planets of 1.05$M_{\oplus}$ and 1.25$M_{\oplus}$ and four super-Earths with masses between 3.40$M_{\oplus}$ and 6.60$M_{\oplus}$. It is worth noting that the two Earth-mass planets are the inner planets produced in the HZ in those systems that formed two planets in that region (SIMs 5 and 6).   

{\bf Water delivery:} The planets that survived in the HZ in this dynamical scenario can be classified into two different categories according to the initial location of their accretion seeds.
\begin{itemize}
\item {\it Water worlds:} They are represented by the HZ planets formed in SIMs 2, 3, and 4, the two HZ planets produced in SIM 5, and the outer planet that survived in the HZ in SIM 6. These planets evolved from accretion seeds located beyond the snow line at the end of the gaseous phase. Thus, they had 50\% of water by mass as primordial content. Like in scenario IV, these planets accreted a significant amount of water by impacts. Between 30\% and 50\% of their final water content was provided by collisions with water-rich bodies. In general terms, the water provided by planetesimals is comparable to the water provided by planetary embryos. 
\item {\it Planets in situ:} They are represented by the HZ planet formed in SIM 1 and the inner planet that survived in the HZ in SIM 6. These planets evolved from accretion seeds located inside the snow line at the end of the gas phase. Thus, these planets did not have primordial water, since they were formed in a dry environment. From this, the final water content of these planets was entirely accreted by impacts with water-rich bodies. On the one hand, the planet of interest formed in SIM 1 finished the simulation as a super-Earth with a final mass of 3.40$M_{\oplus}$ and 15\% of water by mass. This planet received 70\% of its total water content in impacts with embryos and 30\% in impacts with planetesimals. On the other hand, the inner planet formed in the HZ in SIM 6 acquired water only through impacts with planetesimals, and it completed the simulation with a final mass of 1.25$M_{\oplus}$ and 6\% of water by mass. We remark that this planet is the most similar to our home planet Earth formed in the present work, even though its water content is 268 times higher than the amount of water on Earth's surface.\\
\end{itemize}          

The results described in this section allow us to infer that the formation of planets in the HZ is a very efficient process in this dynamical scenario. All $N$-body simulations produced planets in the HZ. Moreover, it is worth noting that four of six simulations formed only one HZ planet, while two of six simulations produced two planets in the HZ. We also remark that six of eight HZ planets were water worlds, while the other two were planets in situ with water accreted by impacts. Like in the previous scenarios, the systems harboring a 2.5$M_{\oplus}$ super-Earth as main perturber were not able to form planets with physical and dynamical properties similar to the Earth. However, this scenario was capable of producing an Earth-mass planet, but with a water content 268 times higher than that on the Earth's surface.

\section{Conclusions and discussions}

We carried out a study aimed at analyzing the formation of planets in the HZ and water delivery around solar-type stars in different dynamical scenarios. These scenarios were defined from the most massive planet of the system, which was formed just beyond the snow line at 3 au at the end of the gaseous phase. According to this, we focused our research on five scenarios, each of which harbored a planet analog to Jupiter (I), Saturn (II), Neptune (III), or a super-Earth of 5$M_{\oplus}$ (IV) and 2.5$M_{\oplus}$ (V) around the snow line. 

This work allowed us to derive several interesting conclusions related to the formation and evolution of planetary systems around solar-like stars.

\begin{itemize}
\item {\it The formation of planets in the HZ seems to be a common process around Sun-like stars.} Our results show that the most common kind of planets formed in the HZ are water worlds, that
is, terrestrial-like planets with very high water contents. Most of the planets formed in the HZ in the present work are water worlds, with the exception of three dry worlds and one Earth-mass world with 6\% of water by mass. This would suggest that the formation of planets in the HZ with masses and water contents similar to the Earth around Sun-like stars seems to be a rare process. 

\item {\it The process of formation of planets in the HZ is still
efficient in extreme dynamical environments.} On the one hand, our results show that planets are efficiently produced in the HZ in systems harboring a super-Earth around the snow line at the end of the gaseous phase. All but one of the numerical simulations carried out in such a dynamical scenario formed at least a planet in the HZ with water content. On the other hand, 10 of 15 numerical simulations that harbored a Jupiter-like planet around the snow line at the end of the gaseous phase produced a planet in the HZ, while in 7 of these 10, the planets that formed had water. These results could have important implications in the search of extrasolar planets around Sun-like stars. It could be promising for exoplanet observation programs with astrobiological interest to select planetary systems of interest to observe. 

\item The water content of the planets surviving in the HZ has two possible origins: primordial (water accreted from the protoplanetary disk during the gaseous phase), and collisional (water accreted by collisions with water-rich embryos and planetesimals during the post-gas phase). In general terms, a large fraction of the planet's water content is primordial. We remark that all water content of the HZ planets in systems that harbor a Jupiter-mass planet around the snow line is primordial. However, in planetary systems with less massive planets acting as main perturbers, the collisional water contribution gains relevance. Collisional water contribution increases as the mass of the most massive planet of the system decreases.
\end{itemize}

However, some careful considerations must be made regarding these results. The MERCURY code used to carry out our $N$-body simulations treats all collisions as inelastic mergers that conserve the total mass and the water content of the system of interacting bodies. Thus, the masses and water contents of the planets formed in the HZ in this study should be interpreted as upper limits. \citet{Chambers2013} developed $N$-body simulations including fragmentation and hit-and-run collisions rather than assuming that all impacts lead to a perfect merger of the colliding bodies. This improved model is based on the results of hydrodynamical simulations of planetary impacts performed by \citet{Leinhardt2012} and \citet{Genda2012}, who identified the boundaries of different collisional regimes and provided formulae for the mass of the largest remnant. Based on this, \citet{Chambers2013} carried out $N$-body simulations of terrestrial planet formation incorporating collisional fragmentation and hit-and-run collisions, and he
then compared them to simulations in which all collisions were assumed to result in mergers. The final planetary systems formed in the two models were broadly similar. However, the author found differences concerning the planetary masses and the time-averaged eccentricities of the final planets. On the other hand, \citet{Dvorak2015} developed hydrodynamic simulations to quantify the water retained in fragments after a collision for different velocities and impact angles. The authors found that most of the water remains on the survivor for impact angles $\alpha \lesssim 20^{\circ}$ and velocities $v \lesssim 1.3 v_{\text{esc}}$, while more than 80\% of the water also remains in strongly inclined hit-and-run scenarios. In general, an increasing amount of water is lost in debris for small impact angles and high velocities. \citet{Dvorak2015} observed significant water loss of up to 60\% for faster and/or less inclined impacts. Based on these considerations, future $N$-body simulations should include a more realistic treatment of the collisions and the evolution of water in order to determine the orbital and physical properties of the planets in more detail.

Our investigation has improved our understanding of the formation and evolution of habitable planets around solar-type stars. Future observational evidence is needed to test our theoretical models, while future works will allow us to answer more questions about the nature of the different worlds in the Universe. 

\begin{acknowledgements}
This work was partially financed by CONICET by grant PIP 0436/13.
\end{acknowledgements}

\bibliographystyle{aa} 
\bibliography{30848} 

\end{document}